\documentclass[twocolumn]{aastex631}

\usepackage{color}
\usepackage{hyperref}
\usepackage{multirow}
\usepackage[normalem]{ulem}             

\renewcommand{\ion}[2]{#1\,{\sc #2}}

\newcommand{\chianti}{\textsc{chianti}}

\shorttitle{Advanced ionization equilibrium models}
\shortauthors{Dufresne et al.}
\graphicspath{{./}{*}}

\begin{document}

\title{CHIANTI -- an atomic database for emission lines - Paper XVIII.  Version 11, advanced ionization equilibrium models: density and charge transfer effects. }


\author[0000-0002-6946-4722]{R. P. Dufresne}
\affiliation{DAMTP, Center for Mathematical Sciences, University of Cambridge, Wilberforce Road, Cambridge, CB3 0WA, UK}

\author[0000-0002-4125-0204]{G. Del Zanna}
\affiliation{DAMTP, Center for Mathematical Sciences, University of Cambridge, Wilberforce Road, Cambridge, CB3 0WA, UK}

\author[0000-0001-9034-2925]{P. R. Young}
\affiliation{NASA Goddard Space Flight Center, Code 671,
  Greenbelt, MD 20771, USA}
\affiliation{Northumbria University, Newcastle Upon Tyne NE1 8ST, UK}

\author[0000-0003-1628-6261]{K. P. Dere }
\affiliation{Department of Physics and Astronomy, George Mason University, 4400 University Drive, Fairfax, VA 22030, USA}

\author[0009-0000-0209-4387]{E. Deliporanidou}
\affiliation{DAMTP, Center for Mathematical Sciences, University of Cambridge, Wilberforce Road, Cambridge, CB3 0WA, UK}

\author[0000-0001-9642-6089]{W. T. Barnes}
\affiliation{NASA Goddard Space Flight Center, Code 671,
  Greenbelt, MD 20771, USA}
\affiliation{Department of Physics, American University, Washington, DC, USA} 

\author[0000-0002-9325-9884]{E. Landi}
\affiliation{Department of Climate, Space Sciences and Engineering, University of Michigan, Ann Arbor, MI, 48109}

\begin{abstract}
Version 11 of the \chianti\ database and software package is presented. Advanced ionization equilibrium models have been added for low charge states of seven elements (C, N, O, Ne, Mg, Si and S), and represent a significant improvement especially when modelling the solar transition region. The models include the effects of higher electron density and charge transfer on ionization and recombination rates.
As an illustration of the difference these models make, a synthetic spectrum is calculated for an electron pressure of 7$\times 10^{15}$\,cm$^{-3}$\,K and compared with an active region observation from HRTS. Increases are seen of factors of two to five in the predicted radiances of the strongest lines in the UV from \ion{Si}{iv}, \ion{C}{iv}, and \ion{N}{v}, compared to the previous modelling using the coronal approximation. Much better agreement (within 20\%) with the observation is found for the majority of the lines. The new atomic models better equip both those who are studying the transition region and those who are interpreting emission from higher density astrophysical and laboratory plasma. In addition to the advanced models, several ion datasets have been added or updated, and data for the radiative recombination energy loss rate have been updated. 
\end{abstract}


\section{Introduction} \label{sec:intro}

The \chianti\ atomic database \citep{dere1997,dere2023} is widely used in astrophysics to model line emission across the wavelength spectrum, but it is also increasingly used to model laboratory, high density plasma \citep[see][for example]{trabert2022,kambara2021}. Significant effort has been devoted over numerous releases to improve the atomic rates used to calculate the emissivities of the spectral lines emitted by the most important ions for astrophysics.
\chianti\ is now the most widely-used atomic database for solar physics, and has become, in certain cases, the reference for other atomic databases in astrophysics. For example the \textsc{cloudy} spectral synthesis code was recently updated \citep{chatzikos2023} to include the \chianti\ 10.0.1 atomic data.

The main aim of this \chianti\ release is to improve the ionization equilibrium calculations for complex ions forming in higher density plasma. One principal assumption within \chianti\ from the beginning has been that the plasma is in ionization equilibrium, and the ion abundances have been pre-calculated assuming the so-called `coronal approximation' \citep[e.g. Eq. 24 of][]{delzanna2018}. This assumes that, for ionization and recombination purposes, atoms and ions are entirely populated in their ground state. Many of the other key assumptions in the coronal approximation are suitable only for high temperature, low density plasma; 
however, even at typical electron densities of the quiet-Sun corona some of the assumptions break down. The problem is magnified further in the higher density transition region (TR). Various effects have been invoked in the atomic modelling over the years to improve the agreement between observations and theory for the TR, as discussed below.

The current missions observing the solar TR, the Interface Region Imaging Spectrometer \citep[IRIS][]{depontieu2014} and the Spectral Imaging of the Coronal Environment \citep[SPICE][]{anderson2020} spectrometer on Solar Orbiter, observe many lines from \ion{C}{ii}, \ion{O}{ii}, \ion{O}{iv}, \ion{O}{vi}, \ion{Si}{iv} and \ion{S}{iv}, for example. These are used for a wide range of plasma diagnostics, but all show discrepancies between observations and theory. Although often not a physical representation for the plasma emission, since the 1960's various emission measure techniques were applied to observations. With the above simple assumption of ionization equilibrium and the coronal approximation, it was found that much of the plasma emission could be well represented, except several so called `anomalous' ions from the  Li- and Na-like sequences, such as \ion{C}{iv}, \ion{O}{vi}, \ion{Si}{iv}. These are under-predicted by typically a factor of 5 compared to those from other ion sequences which form at similar temperatures \citep[see,e.g.][]{burton1971,dupree1972}.

The main anomalous ions, such as \ion{Si}{iv}, \ion{C}{iv}, \ion{N}{v}, and \ion{O}{vi}, emit some of the strongest lines in the UV and the discrepancies have limited their potential use for plasma diagnostics. For example, to use the \ion{Si}{iv} 1402.77\,\AA\ to \ion{O}{iv} 1401.16\,\AA\ ratio for density diagnostics in flares from IRIS, \citet{young2018iris} applied an empirical correction factor of 3 to the \ion{Si}{iv} line intensity. 
\citet{doschek2001} noted that the \ion{Si}{iv} and \ion{O}{iv} lines form at different temperatures ($6.3\times10^4\,$K and $1.6\times10^5\,$K respectively, using the coronal approximation) which makes it difficult to derive diagnostics from such lines. \citet{young2018iris} and \citet{doschek2001} both used \chianti\ for their modelling. \cite{young2018iris} included a modification of the coronal approximation by applying an estimate of the suppression of dielectronic recombination (DR) at higher densities. This is a physical effect first noted by \citet{burgess1969} and was included in the ion balances of \citet{jordan1969} and \citet{summers1974}, for instance. 

Another improvement for the atomic models of Si and O was found by including charge transfer (CT). Charge transfer occurs during atom-ion or ion-ion collisions when an electron is exchanged between the colliders.
Although O is usually only affected by this process in the solar chromosphere, \citet{baliunas1980} showed how all the Si ions forming in the solar transition region are affected. As an example, they estimated that the formation temperature of the \ion{Si}{iii} intercombination line at 1892.03\,\AA\ is 20,000\,K when charge transfer is included, compared to 32,000\,K predicted by the coronal approximation. 

The problem of the anomalous ions is present also in stellar atmospheres, as shown by \cite{delzanna2002}, where EUV and UV observations from several satellites were combined. \citet{sim2005} included an approximate treatment of DR suppression and CT for some ions in their semi-empirical atmospheric models for $\epsilon$ ERI (K2~V), apparently resolving the main discrepancies.

One further improvement used in the atomic modelling is to take account of ionisation and recombination from metastable levels. These longer-lived levels become populated in higher density plasma, which alters the ionization and recombination rates out of the ion. The importance of this was first demonstrated by \citet{nussbaumer1975}, who showed that all the carbon ions in the transition region are formed at lower temperatures than in the coronal approximation. They showed the effect on temperature and density diagnostics for a range of solar conditions when using the \ion{C}{ii} 1334.53\,\AA\ line, which is also observed by IRIS, but note that ions of all elements forming in the transition region should be affected by this process. An approximate treatment of both DR suppression and ionisation from metastable levels plus charge transfer was developed by J.~Raymond, as briefly described in \cite{vernazza_raymond:1979}. In a follow-up paper, \cite{raymond_doyle:1981b} applied a differential emission measure (DEM) method to Skylab observations of the quiet Sun, apparently resolving the main discrepancies, without invoking time-dependent ionization or other effects. 

All of the above-described improvements to the atomic modelling were integrated recently, with updated atomic rates, into models for the main elements observed in the solar transition region (C, N, O, Ne, Mg, Si and S), as described in \citet{dufresne2019,dufresne2020,dufresne2021pico,dufresne2021picrm}. Significant changes were observed in the ion balances at typical densities from the quiet Sun through to flares, especially below $10^5$\,K.
A comparison of these models with results from the coronal approximation and with a compilation of averaged intensities of the quiet Sun \citep{dufresne2023piobs} showed significant improvements.
Changes in predicted intensities by factors of two were shown in a number of cases, such as the \ion{O}{ii} 718.49\,\AA\ lines observed by SPICE. Intensities of the \ion{Si}{iv} resonance lines were enhanced by a factor of six, bringing the ratio with the \ion{O}{iv} intercombination lines much closer to observations. However, some discrepancies with observations for the anomalous ions were still present.

Of course, changes to the atomic modelling are not the only effects that could bring improved agreement. While atomic modelling is known to particularly affect Li- and Na-like ions, an investigation by \citet{judge1995} considered this insufficient to entirely account for the discrepancy in these ions. \citet{judge1995} and other authors have suggested that the main cause of the discrepancy could be the assumption of ionization equilibrium \citep[such as][]{pietarila2004,olluri2013}. Time-dependent ionization (TDI) is at play whenever the timescales for ionization and recombination of an ion are longer than those of the processes affecting the plasma state. Also, photo-ionization can affect significantly the charge states, depending on the ion and  conditions. Atoms and singly or doubly ionized ions can be affected by opacity and a full radiative transfer might need to be performed \citep[such as][]{rathore2015}. Finally, \cite{dudik_etal:2014_o_4} pointed out instead that non-Maxwellian electron distributions could also explain the \ion{Si}{iv} to \ion{O}{iv} anomaly in the IRIS spectra.

The remainder of this paper describes the improvements made to the atomic modelling in \chianti\ for the current release. The next section of this article describes in more detail the atomic processes being included and the various approximations used to incorporate them. Section~\ref{sec:obs} gives examples of how the new models compare with the coronal approximation. It presents a comparison with observations from HRTS of an active region using DEM modelling to show the improvements over the coronal approximation. A section describing other updates to the database follows in Sect.~\ref{sec:otherdata}. A short conclusion is given in the end.

\section{Advanced models for medium- to high-density plasma}
\label{sec:adv}

In this Section we describe the atomic processes and methods used for the new advanced models. The implementation is simpler than that used by  \citet{dufresne2019} and \citet{dufresne2020,dufresne2021pico,dufresne2021picrm} and only the ionization fractions are affected in comparison to the previous \chianti\ models. The new methods do not impact level populations. The models are currently only available for ions of C, N, O, Ne, Mg, Si and S. The new models are switched on by default in the \chianti\ IDL software using the keyword `\texttt{advanced\_model}', but they can be switched off if preferred.
The advanced models are not currently implemented in ChiantiPy.

One key assumption we retain in the present modelling is that the timescales for ionization/recombination, which are related to the electron density, are such that the plasma is in ionization equilibrium. The models included in the present version of \chianti\ use the rates as described in the models referenced above. We provide now a brief description.

\subsection{Collisional ionization}
\label{sec:ci}

Atomic energy levels which have no rapid, dipole-allowed decays to lower energy levels can become significantly populated in certain conditions. For instance, in high electron densities they become populated through the balance of collisional excitation and de-excitation. Such energy levels are denoted as metastable levels.
Rate coefficients for collisional ionization (CI) from metastable levels are generally larger than those from the ground level in the same ion because metastable levels are closer to the continuum. The overall ionization rate coefficient out of the ion is the sum of the total ionization rate coefficient from each initial level weighted by the relative population of the level. When metastable levels become populated in higher density plasma the overall ionization rate coefficient is higher than in a low density plasma. This causes the ions to form at lower temperature. The coronal approximation only includes ionization from the ground level, and cannot take account of this effect.

Collisional ionization rates for ground and metastable states of carbon and oxygen were calculated in \citet{dufresne2019} and \citet{dufresne2020}, respectively. \textsc{Flexible Atomic Code} \citep[FAC,][]{gu2008} was used for direct collisional ionization and \textsc{Autostructure} \citep{badnell2011} for excitation--auto-ionization (EA, or indirect collisional ionization). The cross sections were benchmarked against many of the same experiments as \citet{dere2007}, whose ground-level CI rates have been used in \chianti\ for the default ion balances until now. To provide a consistent set of rates for the advanced models, the rate coefficients from \citet{dufresne2019} and \citet{dufresne2020} are used for ground and metastable levels of carbon and oxygen. 

For other ions included in the advanced models the same method is used as \citet{dufresne2021picrm}. The rate coefficients of \citet{dere2007} are used for the ground levels. To estimate rates for metastable levels, the \citet{burgess1983} CI approximation for low charge ions is used to calculate the ratio of the metastable to ground rate coefficients. The \citet{dere2007} rate coefficients are multiplied by this ratio to estimate those for the metastable levels. The ionization potentials required by the \citet{burgess1983} approximation are taken from experimental values stored in \chianti\ for each ion. \citet{dufresne2021picrm} compared the oxygen ion balance obtained using this approximation with that using \textit{ab initio} CI rate coefficients; differences in the ion balance were negligible.

\subsubsection{Calculating overall ionization rates for neutrals}

To obtain overall ionization and recombination rates for the models requires knowing the level populations within each ion (see Sect.~\ref{sec:solution} for more details). Radiative decay and electron impact excitation rates are needed to determine this. 
\chianti\ 10.1 had rates for all ions of C, N, O and S, but for Ne, Mg and Si \chianti\ did not have the neutral atom. So, it has been necessary to create new models. The \ion{Ne}{i} model represents a new addition to the regular database. Only approximate models could be constructed for \ion{Mg}{i} and \ion{Si}{i} which are not suitable for computing accurate line emission. Therefore, although the data files have been added to the database, the ion names have been omitted from the \chianti\ `masterlist' file for the latter two ions. (The \chianti\ User Guide provides more detail on how ions in the `masterlist' are treated.)

For \ion{Ne}{i} energies and radiative decay rates are taken from the energy-adjusted calculation of \citet{froese2004}, as made available on the NIST MCHF collection website\footnote{https://nlte.nist.gov/MCHF/}. Excitation data are taken from the B-spline R-Matrix calculation of \cite{zatsarinny2012} in \textit{jK}-coupling, and were made available by Professor Bartschat (2024, private communication). Since all the data are obtained from \textit{ab initio} calculations the data are suitable for spectroscopic analysis of this ion.

All data in \chianti\ are resolved by fine structure, but the only available excitation data for \ion{Mg}{i}, from \citet{barklem2017}, are in \textit{LS}-coupling.
In cases where resolution by fine structure is required in the data, transitions involving multiplets are often split according to the statistical weights of the levels. This approach is very approximate and relevant only for plasma in which densities are high enough that energy levels within a term are populated according to statistical weight. Another way of splitting the data was tested here. \textsc{Autostructure} (AS) was used to calculate excitation data for \ion{Mg}{i} in intermediate-coupling. The \textit{LS}-coupling data of \citet{barklem2017} was split into the same ratios as the AS intermediate-coupling data for the levels within each term. This second approach was found to be preferable because it takes into account the relative strengths of the transitions within a multiplet seen in the intermediate-coupling data, rather than distributing them purely by statistical weight. (AS cannot be used in itself for the neutral excitation data because it implements the distorted wave method, which is known to be insufficiently accurate compared to non-perturbative methods for low charge ions.)

As a result of splitting the rates, the data for \ion{Mg}{i} should be used only for calculating metastable ionization and recombination, which is not as sensitive to accuracy in level populations as high resolution spectroscopy. There are two types of excitation data in \citet{barklem2017}; the B-spline R-Matrix excitation data were used here, as provided by Professor Bartschat (2024, private communication). Again, level energies and radiative data were taken from \citet{froese2004}, but this time using the \textit{ab initio}, not energy-adjusted, data because more transitions were included.

The same simple model for \ion{Si}{i} is used as in \citet{dufresne2021picrm}. This used energy levels and radiative data from \citet{fischer2005}, supplemented by additional radiative data from a calculation using ATSP2K \citep{fischer2007atsp}. As above, since the excitation data came from a distorted wave calculation, it is strongly recommended that the data for this ion are not used for any purpose other than calculating overall ionization and recombination rates.

\subsection{Radiative and dielectronic recombination} \label{sec:rrdr}

Since metastable levels are further away in energy than the ground from levels in the next lower charge state, rate coefficients for radiative recombination (RR) and dielectronic recombination (DR) are generally smaller for metastable levels than the ground. This will also cause ions to form at lower temperatures when metastable levels become populated compared to the coronal approximation.

We use the RR rate coefficients calculated by \citet{badnell2006}. The DR rate coefficients for ground and metastable levels have so far been calculated for all ions of hydrogen to zinc in the H-like to P-like isoelectronic sequences by the DR Project \citep[see][for the first paper in the series]{badnell2003}. Rate coefficients were calculated for all ground and metastable levels up to the first dipole allowed transition in the ion. 
Previous versions of \chianti\ only included the RR and DR rate coefficients for ground levels, using the fitting coefficients made available by N.R. Badnell\footnote{http://apap-network.org/}. We now include in \chianti\ the RR and DR rates for metastable levels using the same set of data. Rate coefficients were calculated to final resolved states at zero density in those works; here, we use total rate coefficients resolved by initial level to the next lower charge state.

\subsubsection{Dielectronic recombination suppression} \label{sec:drsuppr}

DR mostly goes through highly excited, Rydberg levels close to the continuum. In higher density plasma these levels are rapidly ionized by free electrons before decays to lower levels take place. The effects are not easy to calculate, as they require complex collisional-radiative models and a large number of rates which are not easy to calculate accurately. Simple hydrogenic models with very approximate atomic rates were developed by \citet{burgess1969,burgess1976}. The tables of the resulting effective recombination rates obtained by \citet{summers1974} have then been used by the authors previously mentioned \citep{vernazza1979,judge1995,jordan1969} to approximate DR suppression. 

\begin{figure}
 \centering
 \includegraphics[width=8.5cm]{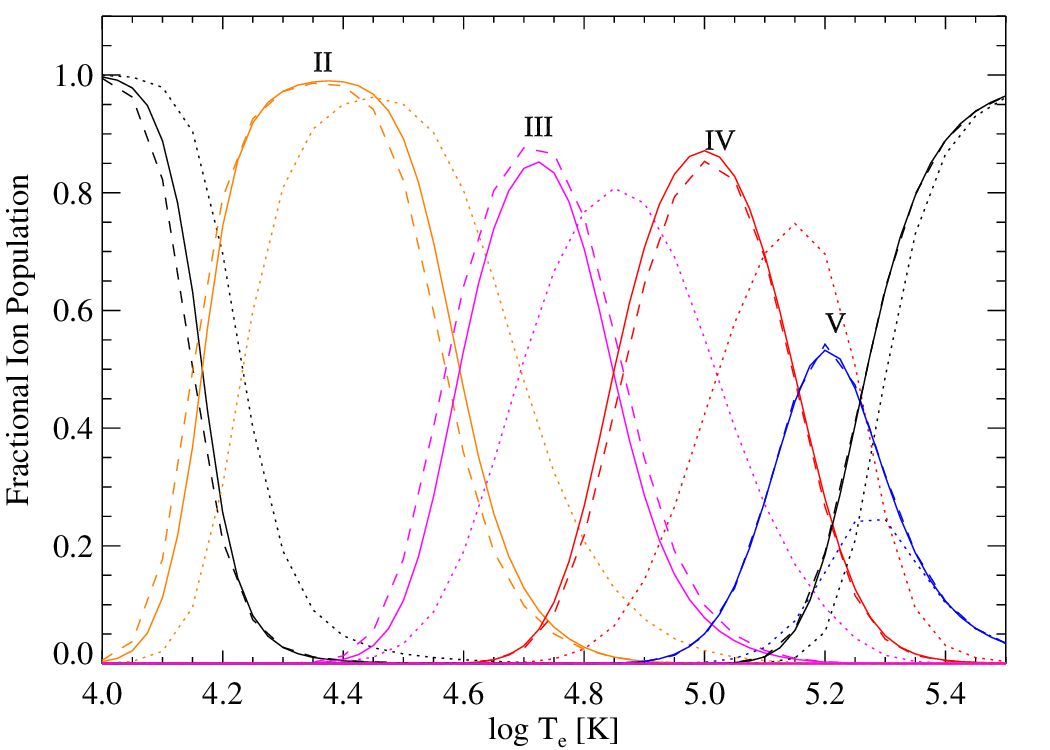}
 \caption{Comparison of the effect of different approximations for DR suppression on the nitrogen, density-dependent ion balance of \citeauthor{dufresne2021picrm} at $10^{12}$~cm$^{-3}$ density: solid line - using DR suppression from \citeauthor{summers1974} tables, dashed - using DR suppression from \citeauthor{nikolic2018}, dotted - coronal approximation. Ions are indicated by Roman numerals for the effective ion charge and by different colours.}
 \label{fig:ncrmsuppr}
\end{figure}

In a recent version of \chianti\ \citep[v10,][]{delzanna2021v10} the \citet{nikolic2018} approximation to reproduce the suppression seen in the \citet{summers1974} rates was introduced. We use the \citet{nikolic2018} approximation in the present advanced models to reduce the DR rates for both the ground and metastable levels. 
Although the same ionization and recombination rates are being used for the present models as \citet{dufresne2021picrm} and the preceding works, there are some differences in the ion balances because the earlier works use DR suppression factors taken from the \citet{summers1974} tables. The \citeauthor{nikolic2018} approximation is used because it is significantly slower to interpolate the \citeauthor{summers1974} data in both temperature and density to obtain suppression factors for the models. Figure~\ref{fig:ncrmsuppr} shows the difference in the N ion balance when using these two approaches. Differences can be seen, but comparing the two approaches with the coronal approximation shows that it is more important to have an estimate of DR suppression than ignore it altogether. Differences between the two methods has only been assessed for elements in the advanced models and it is not switched on for other cases. The user can include it for all elements using the `\texttt{dr\_suppression}' keyword, if desired.

\subsection{Charge transfer and associated model atmospheres}
\label{sec:ct}

Another physical effect that changes significantly the ion balance for some ions in the chromosphere and TR is charge transfer (CT) during collisions with the most abundant species: hydrogen and helium. Charge transfer is the exchange of an electron that takes place between two colliders during atom-ion or ion-ion collisions. The process is variously known as charge transfer, charge exchange and electron capture. \citet{baliunas1980} demonstrated that silicon is strongly affected in the solar atmosphere, but it was only relatively recently that more accurate, quantum-mechanical calculations became available, many of which included rates for metastable levels. Earlier calculations used the rather approximate Landau-Zener method, but its assumptions are more suitable for higher collision energies \citep{bates1962ct}.

\citet{dufresne2021pico} and \citet{dufresne2021picrm} made a comparison of all CT cross sections available for the low charge ions of the elements being modelled. The preference was given for results from more accurate methods and those which included metastable levels. Rate coefficients were calculated from the cross sections if not published in the original articles. All except one calculation was in $LS$-coupling and rate coefficients were split according to statistical weights of the initial levels in each term; ground and metastable levels are populated in these ratios for all the ions under consideration in the solar TR. Double electron capture is also possible and was included in the earlier models, but the ion balances were not affected by this and it has been neglected in \chianti.

The CT rate also depends on the number density in the plasma of the relevant perturber, atomic or ionized hydrogen or helium in this case, and its ion fractions. Self-consistent calculations of these values are not feasible and they are usually taken from model atmosphere calculations. The relevant number densities from the model atmospheres are interpolated in temperature over the temperature grid required for the ion balance. Model atmospheric data can be tabulated and read in for the calculation. A number of files have been prepared for the present version, including those of \citet{avrett2008} and \citet{fontenla2014}; data from the latter include the quiet Sun, an active region, plage and facula.

\subsection{Solving the ion balances}
\label{sec:solution}

One of the main differences here compared to the earlier TR models is the solution of the ion balances. The earlier models included in one large matrix all the rates connecting the metastable levels to the ground states of the lower and higher charge states, plus all the rates required within the each ion. The matrix was then inverted to find the populations of all the levels of all the charge states at once. We adopt here a simplified, faster method.

In \citet{dufresne2021pico} the level populations were solved using models in which CI and CT were fully level-resolved. In densities typical of the solar TR, they found that the level populations were all within 2\% of the level populations from the \chianti\ independent atom model, except one level in \ion{O}{ii} which had a difference of 7\%. This is because, for the main lines under consideration in the advanced models, collision rates between levels in an ion are much faster than processes connecting ions. It means in these conditions total CI and CT rates can be used, that is, rates which are resolved only by initial level and not final level. (\citealt{dufresne2021pico} and the other related models used total RR and DR rates. Level populations can be affected by level-resolved RR and DR, but such models are too large and complex to be included in the \chianti\ models. The issues associated with such models can be found in \citealt{delzanna2020}, for example.)

These conditions allow the independent atom model already implemented in \chianti\ to be exploited. The level populations within each ion are calculated first to find the ground and metastable populations. From these, overall ionization and recombination rates out of the ion can be calculated. For example, if $S_i$ is the ionization rate from level $i$, which has a fractional population $n_i$, then the overall ionization rate out of the ion is

\begin{equation}
    S ~=~ \sum_i ~ n_iS_i ~,
    \label{eqn:effrate}
\end{equation}

\noindent where the sum is over all metastable levels. This replaces $S_g$, the total ionization rate from the ground level, used to solve the coronal-approximation ion balance. 

For neutrals, the \chianti\ routine `\texttt{metastable\_levels}' is used to define the levels for which ionization data is included. In the routine, metastable levels are defined as those levels for which there is no decay rate above 10$^5$\,s$^{-1}$. For ions, however, the only metastable levels included in the overall rates are those for which recombination data has been calculated, (see Sect.~\ref{sec:rrdr} for the criteria). The ions included in the advanced models are given in a new master list. For all other ions it is assumed, as previously, that the population is in the ground state for ionization and recombination purposes, and there is no suppression of DR rates with density. 

Once the overall rates are calculated, the same method previously used in \chianti\ to calculate the ion populations is used. Essentially, the ratio of the populations of two successive charge states is proportional to the ratio of the ionization/recombination rates. We have verified that the large matrix approach and the independent atom model produce the same ion abundances, within a fraction of a percent. 
The ion charge states can be both calculated on-the-fly and stored in \chianti-format files for later use. The IDL software has been modified to calculate the advanced ion models by default, and a program has been provided to compare different ionization equilibria. More details are provided in the software notes.

\section{Examples of ion balances and comparison with solar observations}
\label{sec:obs}

It is well known that the greatest effects on ion balances are found at high electron densities, particularly in the \ion{Si}{iv}, \ion{C}{iv} and \ion{N}{v} lines emitted in the transition region \citep[see][for example]{doyle2005}. To illustrate the effects of the advanced models we create synthetic spectra for an active region (AR) using the new and existing models in \chianti, and then compare them with solar observations of an AR. Throughout, we choose these simple assumptions for the modelling: the plasma is in ionization equilibrium; its temperature distribution can be modelled with a DEM which is a single-valued function of temperature;
the atmosphere is static and filling the volume. Another common assumption for the TR is that of constant electron pressure. The model atmosphere data made available here from radiative transfer calculations with a static atmosphere show almost constant electron pressure through the TR, and so we also use this in the present analysis. The pressure used is 7$\times 10^{15}$\'cm$^{-3}$\,K, as determined from the observations and detailed in Sect.~\ref{sec:methods_obs} below. In the ion balances we use the \citet{fontenla2014} facula model atmosphere for the CT data because its pressure throughout the TR matches the pressure determined from the observations.

We acknowledge that, in reality, the solar transition region is actually highly dynamic, such that time dependent ionization can be important, and the physical structure of the emission is possibly highly filamentary. For this reason, proper modelling of the TR is a complex matter, but the present assumptions are still widely used in the literature and will suffice for the simple comparison provided here. We present the ion balances first to highlight what the new, improved models look like and to help explain the changes they cause in line emission. Following that, we describe the methods used for comparing synthetic line intensities with observations and then present the results.

\subsection{Ion balances}
\label{sec:ion_balances}

\begin{figure*}[!ht]
\centerline{\includegraphics[angle=-90,width=9cm,keepaspectratio]{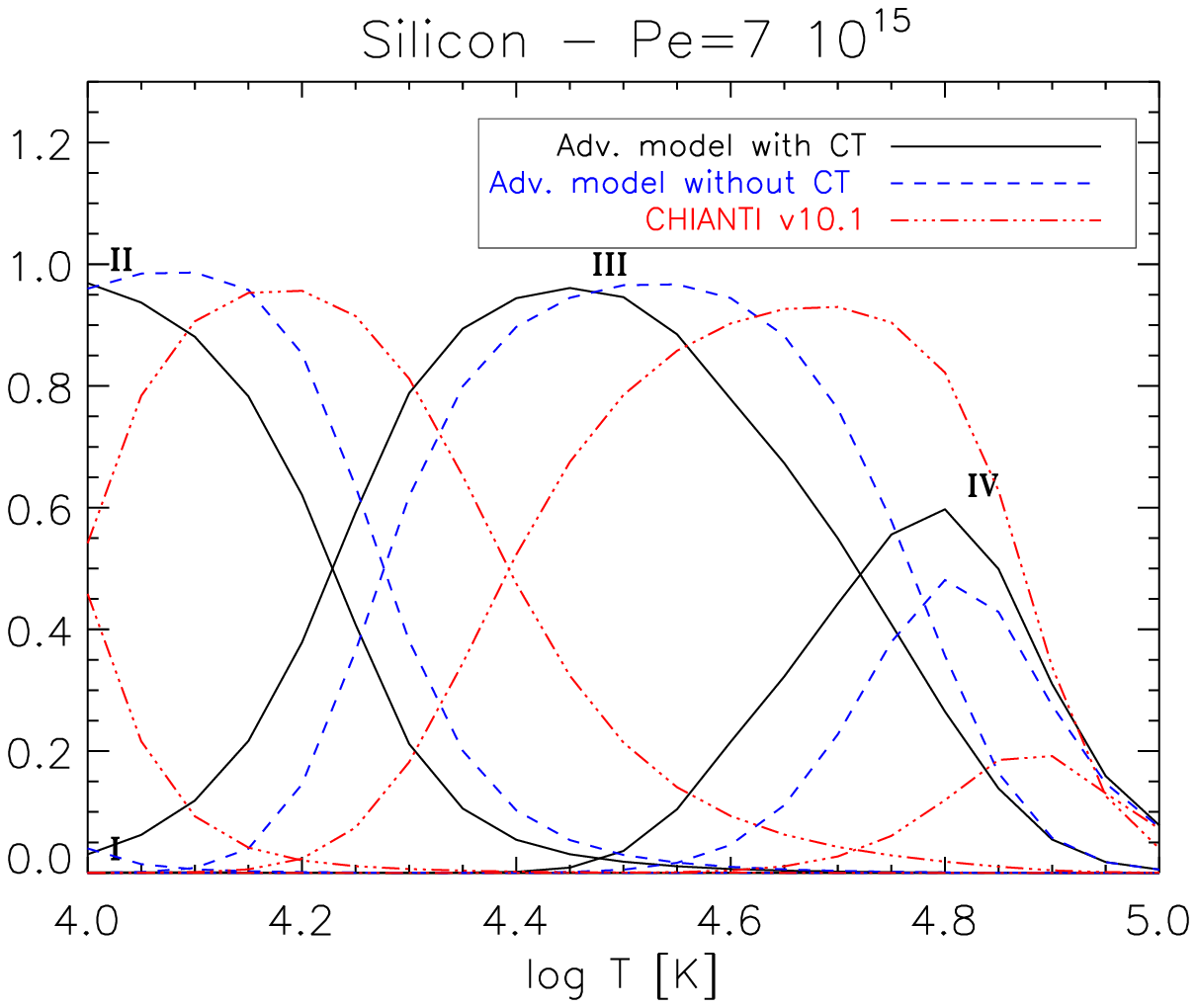}
\includegraphics[angle=-90,width=9cm,keepaspectratio]{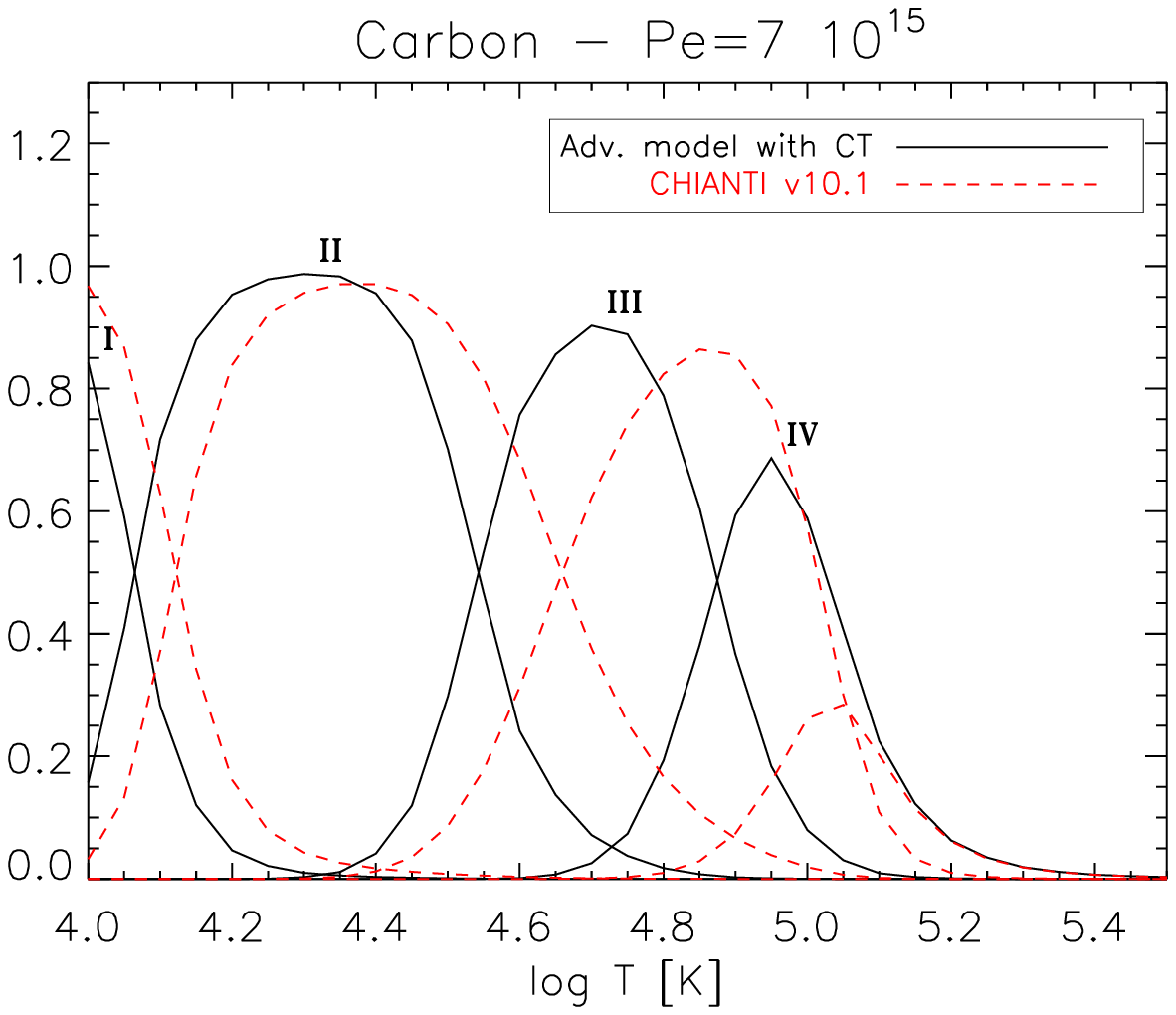}}
\centerline{
\includegraphics[angle=-90,width=9cm,keepaspectratio]{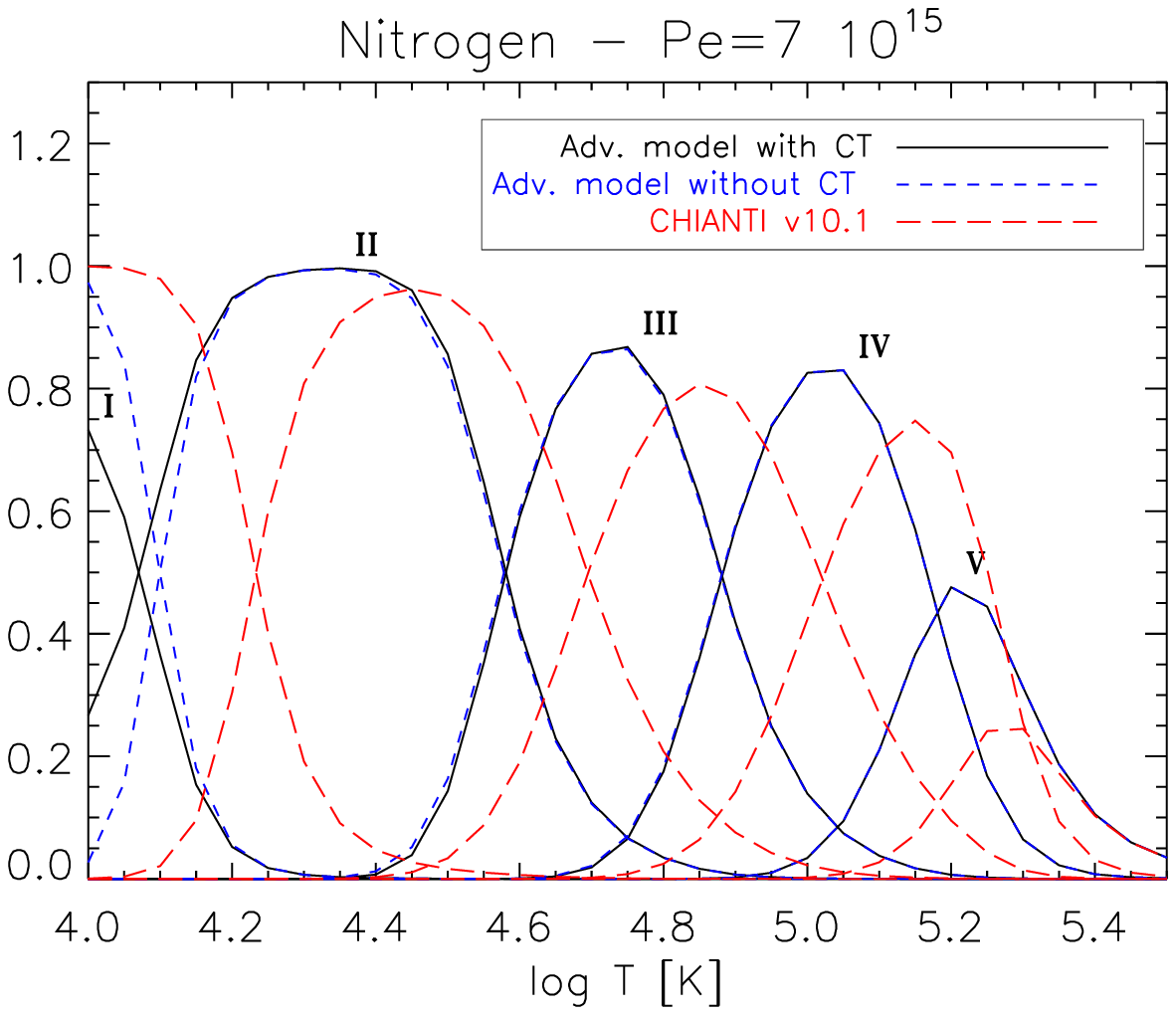}
\includegraphics[angle=-90,width=9cm,keepaspectratio]{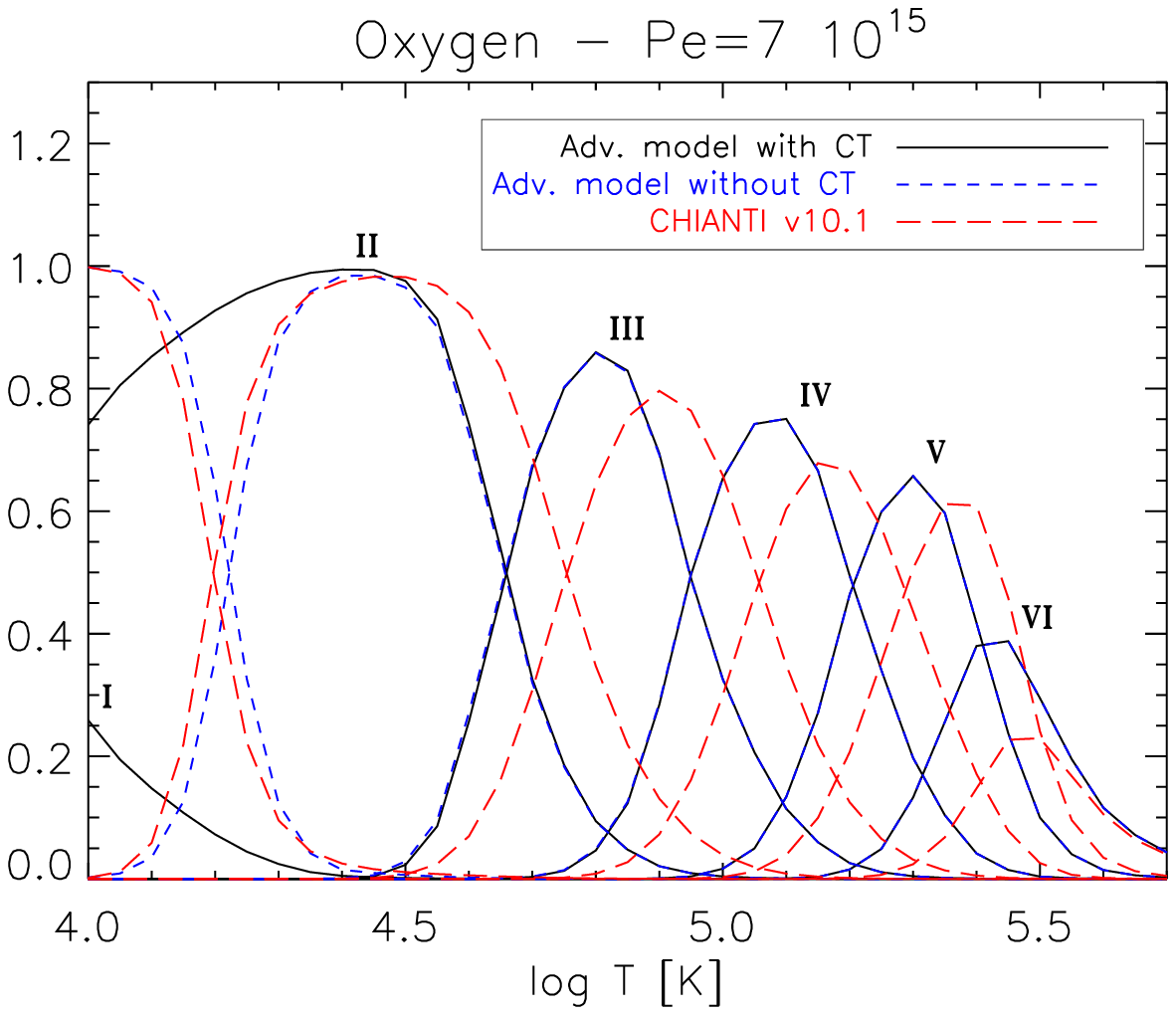}}
\centerline{
\includegraphics[angle=-90,width=9cm,keepaspectratio]{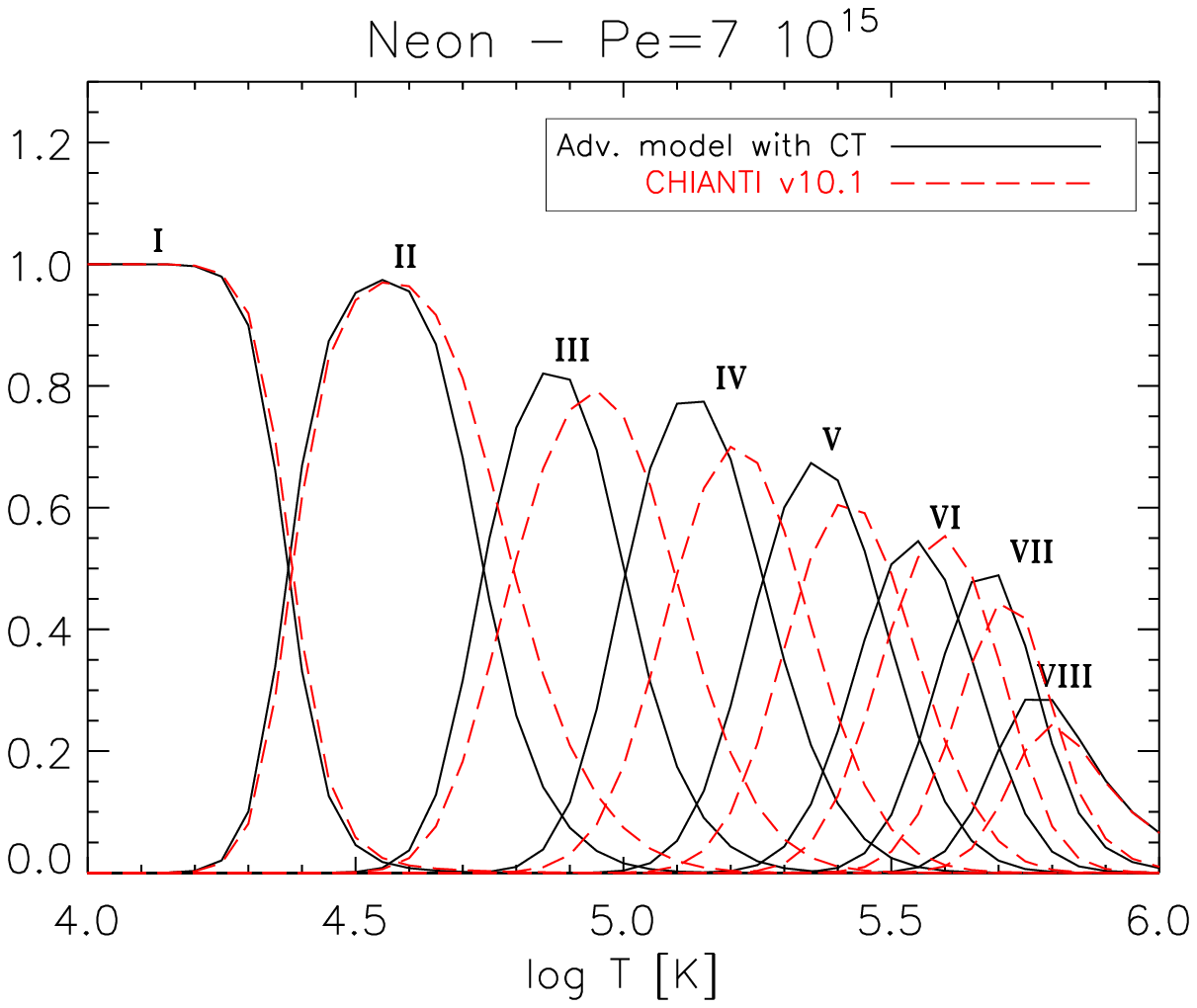}
\includegraphics[angle=-90,width=9cm,keepaspectratio]{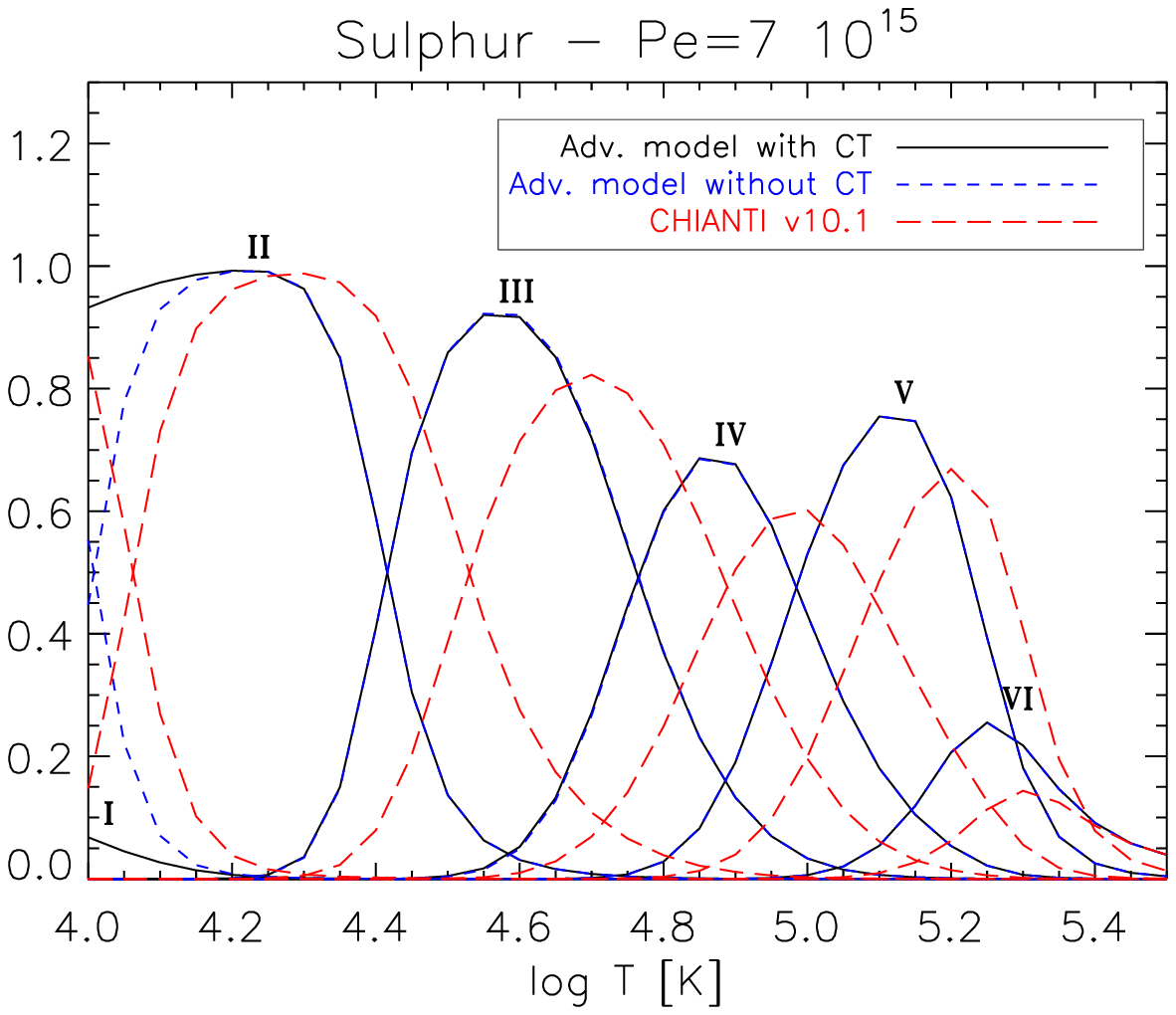}}
\caption{Ion balances calculated with the present advanced models at a constant electron pressure of 7$\times 10^{15}$  cm$^{-3}$  K, with CT included using the \citet{fontenla2014} facula model atmosphere.  We also show the advanced models without charge transfer (with the exceptions of C and Ne for which it has no effect), and the \chianti\ v.10.1 coronal approximation models.}
\label{fig:ion_balances}
\end{figure*}

Figure~\ref{fig:ion_balances} shows the new ion fractions calculated at the selected pressure compared to the zero density \chianti\ v.10.1 models for the Si, C, N, O, Ne and S ions of relevance here. The plots illustrate the fact that TR ion formation is generally shifted towards lower temperatures. This is caused by ionization and recombination from metastable levels in combination with DR suppression. The peaks of the Li- and Na-like, anomalous ions, especially \ion{C}{iv}, \ion{N}{V} and \ion{Si}{iv}, are clearly enhanced. The peak in the ion fractions increase for most other ion sequences, although not to the same extent. The shifts to lower temperature usually cause lines which form at lower temperatures to be enhanced relative to lines in the same ion which form at higher temperatures. Thus, intensity ratios for lines which are emitted by the same ion can also be affected by the ion balances, and not solely by transitions within an ion.

Carbon and neon are unaffected by charge transfer, while it primarily affects nitrogen, oxygen and sulphur in the chromosphere, as highlighted by Fig.~\ref{fig:ion_balances}. By contrast, it is well known from the literature that all TR ions of Si are significantly affected by CT. The ion balance for Si in Fig.~\ref{fig:ion_balances} shows that ion formation shifts to much lower temperature and \ion{Si}{iv} forms over a much wider temperature range.

\begin{figure}[!ht]
\centerline{\includegraphics[angle=-90,width=7.5cm,keepaspectratio]{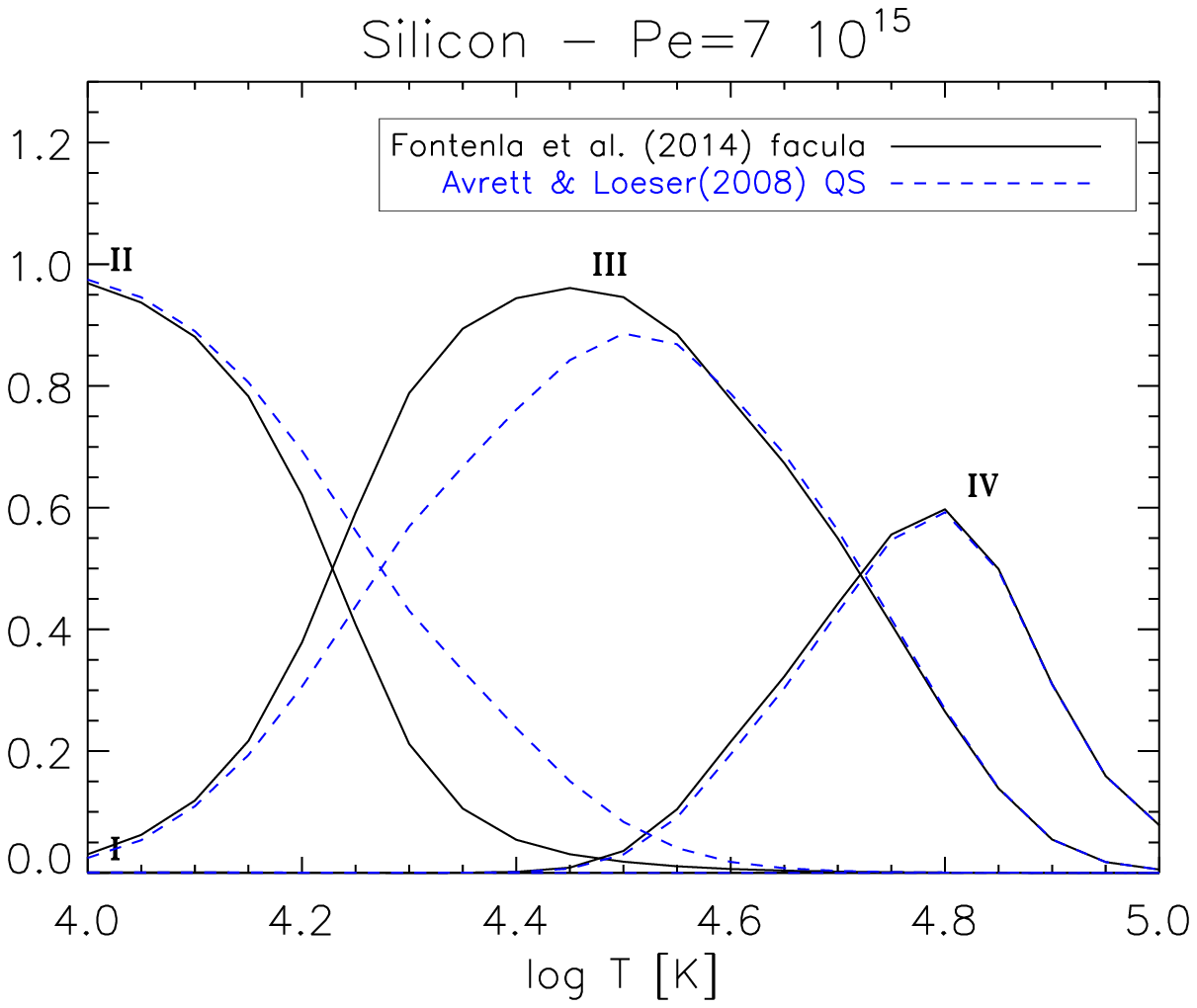}}
\caption{Ion balances calculated with the present advanced models at a constant electron pressure of 7$\times 10^{15}$\,cm$^{-3}$\,K, with CT included, with the \citet{fontenla2014} facula model atmosphere and that of the quiet Sun from \citet{avrett2008}. Note that the main \ion{Si}{iii} lines are formed near log\,$T$=4.6 where differences are small.} 
\label{fig:ion_balances2}
\end{figure}

Inevitably, different atmospheric models will affect the ion balances when CT has an influence because CT rates depend on the total number densities and ion fractions of H and He taken from the model atmospheres. When different model atmospheres within the same work were chosen, such as the quiet Sun (QS) and AR plage from \citet{fontenla2014}, the ion fractions for \ion{Si}{ii} and \ion{Si}{iii} changed by 10\% at the most compared to using the facula model from the same work. However, we show the effects on the ion balances in Fig.~\ref{fig:ion_balances2} that can be obtained when using data from entirely different works. These are an AR facula from \citet{fontenla2014} and the QS model from \citet{avrett2008}. The main effect is caused by CT ionization and recombination between \ion{Si}{ii} and \ion{Si}{iii} during collisions with H.

Collisions with helium affect the ion balance between \ion{Si}{iii} and \ion{Si}{iv}. In both ion balances shown in Fig.~\ref{fig:ion_balances2}, the default He fractions from \chianti\ were used because neither of the model atmospheres provide He data. Despite this, tests using the He model of \citet{delzanna2020}, which in some cases had He ion fractions significantly different than \chianti, produced changes in ion formation of less than 10\%. Even with the differences in the ion fractions caused by the model atmospheres, it is noted that the main \ion{Si}{iii} and \ion{Si}{iv} lines in the HRTS wavelength range have contribution functions which peak at temperatures higher than the peak in the ion abundance. So, the choice of the atmospheric model does not affect the calculated line intensities for these ions, as discussed below.

Neon is not affected by CT, but \citet{dufresne2021picrm} showed that it was the element most affected by photo-ionization, which should be considered when modelling ions up to \ion{Ne}{iv} because it is not currently implemented in \chianti. All the TR ions of Ne are shifted to lower temperature and higher peak ion fraction in the advanced models, although the changes are smaller than those seen for O. Figure~\ref{fig:ion_balances} shows that \ion{S}{i}, like \ion{Si}{i}, is significantly depleted in the chromosphere. For the S ions which form in the TR the shifts to lower temperature and increases in peak ion fractions caused by level-resolved ionization and recombination are reasonably large. Changes to Na-like \ion{S}{vi}, however, are more modest than the other Li- and Na-like ions except for \ion{Ne}{viii}. 

The only element in the advanced models for which the necessary CT data are not available is Mg; this is because there are no CT calculations that include metastable levels for the relevant ions. However, tests were carried out by \citet{dufresne2021picrm} to assess whether the process affects this element. They used the coronal approximation and the rate coefficients from the compilation of \citet{kingdon1996}, and found that \ion{Mg}{i} was almost completely ionized in the solar chromosphere in quiet Sun conditions. This obviously means \ion{Mg}{ii} is the dominant species in this region, although none of the higher charge states are affected by the process. Density dependent effects on free-electron ionization and recombination have been included for Mg, although this affects \ion{Mg}{iv-x} formation to a lesser extent than the changes shown in Fig.~\ref{fig:ion_balances} for the second row elements. Changes to the high charge states (Li- to F-like sequences) of Si and S are also relatively small.

\subsection{Comparison with observed line intensities}

\subsubsection{Observational data used}
\label{sec:methods_obs}

Of the lines from anomalous ions mentioned above, IRIS observes only the \ion{Si}{iv} doublet and does not record enough lines to estimate the temperature distribution of the plasma. Aside from Skylab, SoHO SUMER observed these lines, but not simultaneously. The NRL High Resolution Telescope and Spectrograph (HRTS) was flown several times, observing the entire 1150--1600\,\AA\ region with an excellent resolution of 0.05\,\AA. 
For the present comparison with observations we have chosen the HRTS-II spectra of an on-disk AR plage, provided by \cite{brekke1993}. 

We selected the main lines and measured their radiances with Gaussian fits. Although the profiles of the lines we selected are not self-reversed, the profiles are not exactly Gaussian. This issue combined with the complex conversion from plate density to radiometric calibration means that radiances have a possible uncertainty of the order of 30\%.
The main lines from  \ion{Si}{iv}, \ion{C}{iv}, and \ion{N}{v} do not show any opacity effects when considering the intensity ratios. However, the brightest line in the \ion{C}{ii} multiplet, at 1335.7\,\AA, appears to be partially affected by opacity because the intensity ratio of the lines within the multiplet differ from the optically thin limit.

The main diagnostics to measure the electron density  within the HRTS wavelengths are given by the \ion{O}{iv} lines. The 1399\,\AA\ line is generally very weak, while the 1406\,\AA\ line is blended with \ion{S}{iv}. De-blending the 1406\,\AA\ line is preferable, as discussed by \cite{rao2022}. However, a better line available to HRTS is the 1407\,\AA\ line, in conjunction with the strongest \ion{O}{iv} line at 1401\,\AA. This ratio indicates a density of 5$\times 10^{10}$\,cm$^{-3}$ using the current data in \chianti, equivalent to a pressure of about 7$\times 10^{15}$\,cm$^{-3}$\,K, which is adopted here.

\subsubsection{Methods used to calculate the synthetic spectra}

We used the `\texttt{chianti\_dem}' routine to calculate the DEM using the selection of lines shown in Table~\ref{tab:hrts}. We adopted the MPFIT method, which searches for the best solution having defined a set of DEM spline nodes and uncertainties; the latter are assumed to be 30\% based on the calibration, while the spline nodes were set at log temperatures (in K) of 4.0, 4.3, 4.6, 4.8, 5.0, 5.2, 5.4, 5.5. 
We made three calculations using the same input parameters (set of lines, elemental abundances, spline nodes), and only changed the ion balances in the input. We first ran the advanced model with charge transfer, obtaining good agreement within uncertainties between observed and predicted radiances. We then calculated the DEM using the advanced models without CT, to highlight which lines are most affected by this process. Finally, we ran everything again but used the coronal ion balances from \chianti\ v.10.1. The photospheric elemental abundances of \citet{asplund2021} were used in all three cases. These methods mean that changes in the predicted line intensities can be understood entirely in terms of the changes seen in the ion balances in Sect.~\ref{sec:ion_balances} and the DEMs derived from them.

Gaussian line profiles were used for the model spectra and the widths are the same for each line. Because changes in ion and line formation temperatures are important, we calculate the effective temperature $T_{\rm eff}$ defined by

$$ T_{\rm eff} = { \int G{\left({T}\right)}~
DEM{\left({T}\right)} ~T~dT \over
{\int G{\left({T}\right)}~DEM{\left({T}\right)}~dT} } \quad .
$$

\noindent This is an average temperature more indicative of where each line is formed because the line contribution function, $G(T)$, is weighted by the emission measure. It is noted that $T_{\rm eff}$ is generally different than the peak of the ion balance or the peak of the $G(T)$. 

\begin{figure*}[!ht]
\centerline{\includegraphics[angle=0,width=9cm,keepaspectratio]{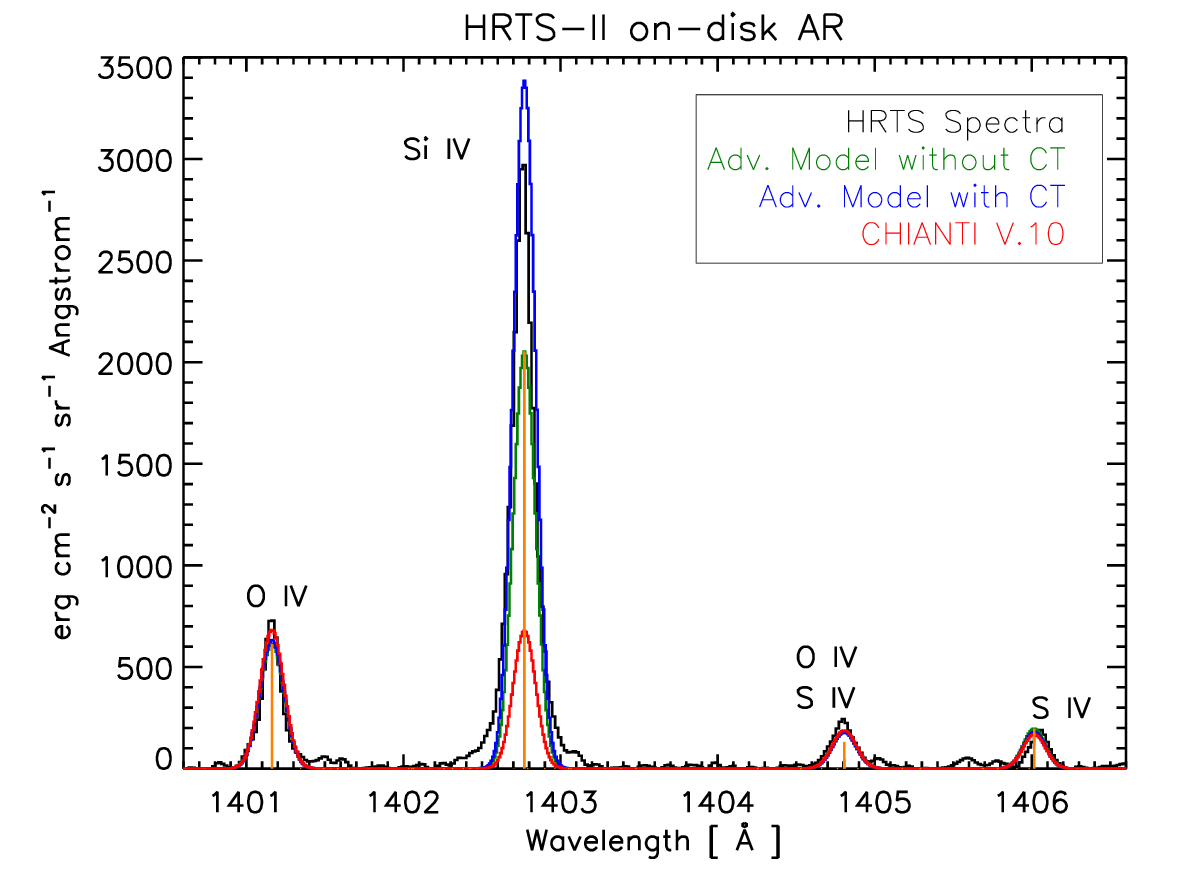}
\includegraphics[angle=0,width=9cm,keepaspectratio]{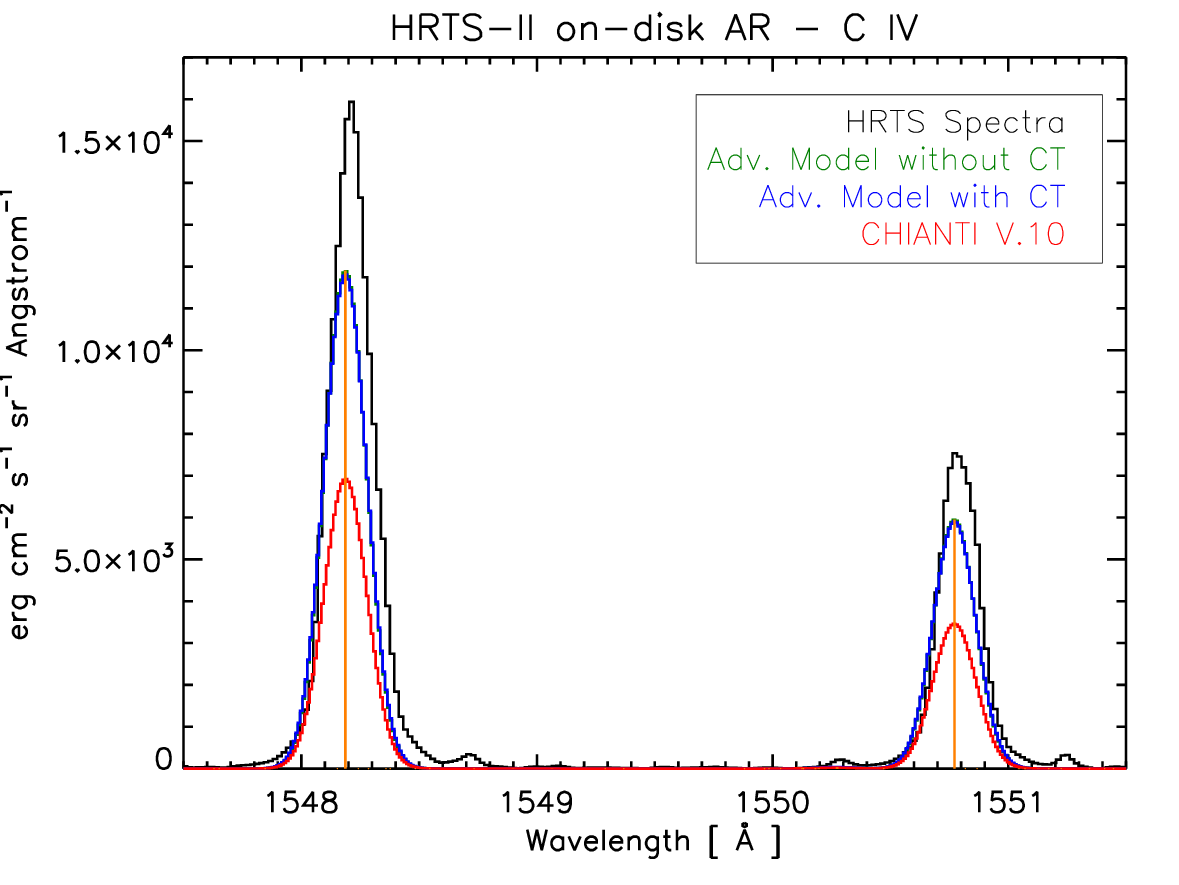}}
\centerline{\includegraphics[angle=0,width=9cm,keepaspectratio]{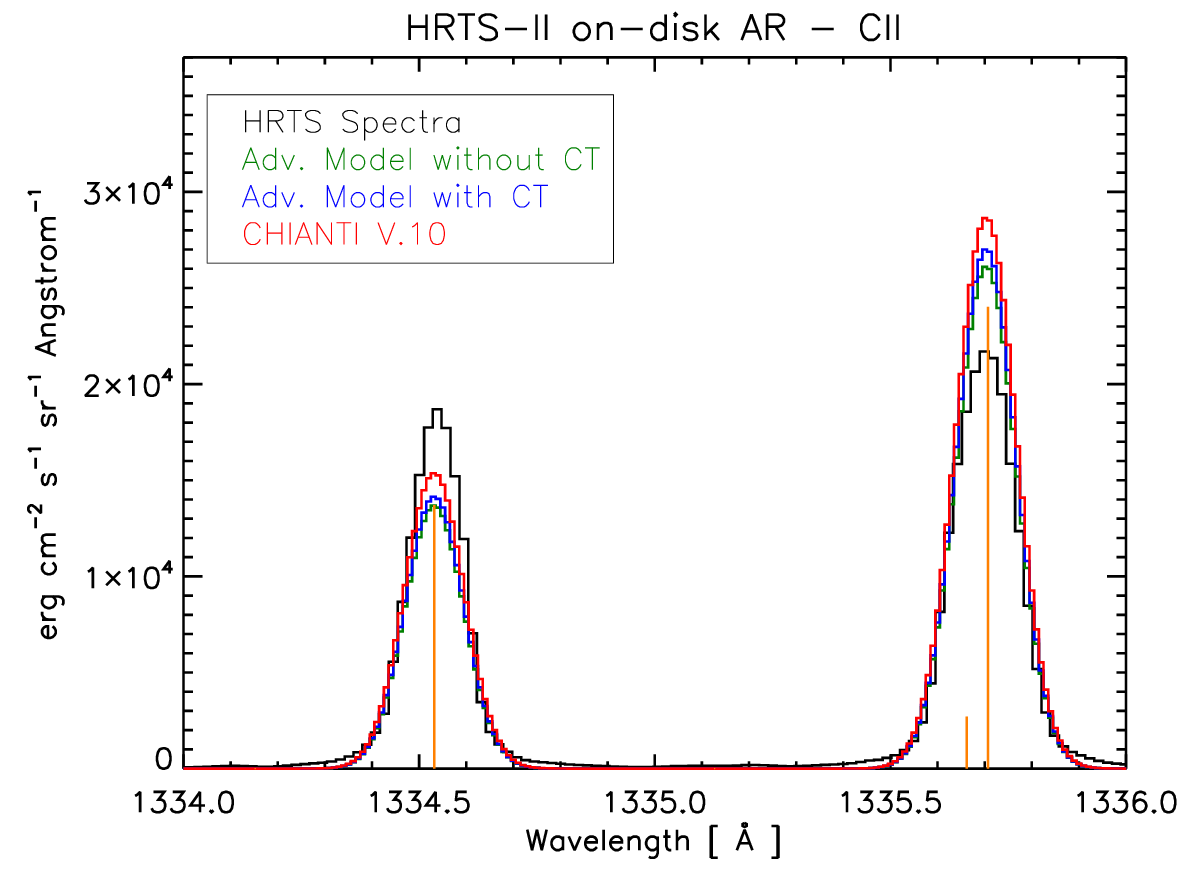}
\includegraphics[angle=0,width=9cm,keepaspectratio]{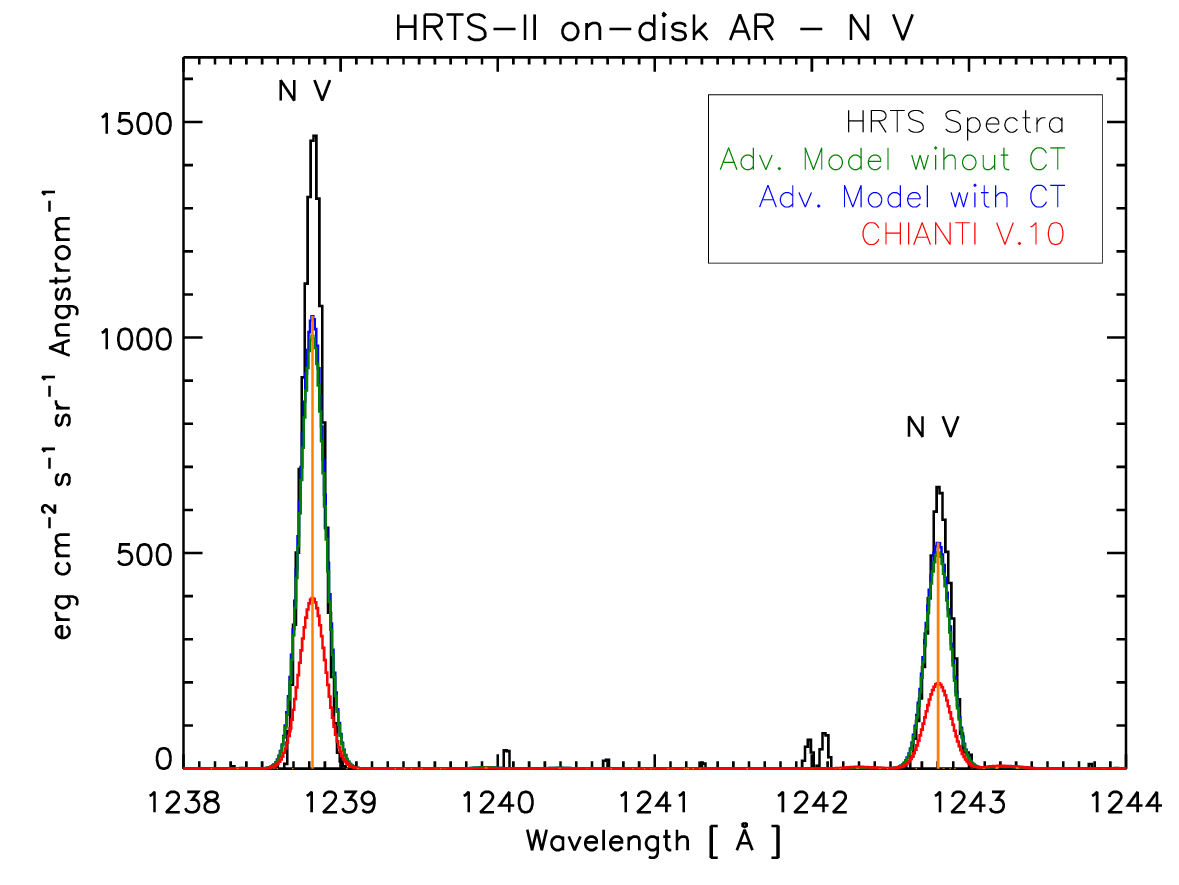}
}
\caption{HRTS-II spectra of an on-disk active region plage, superimposed with various synthetic spectra calculated using the present advanced models at a constant electron pressure of 7$\times 10^{15}$\,cm$^{-3}$\,K. Synthetic spectra obtained with the \chianti\ v.10.1 coronal approximation ion fractions are shown in red, that obtained  from the advanced model without charge transfer is shown in green, and that obtained with the advanced model including charge transfer is shown in blue. The theoretical wavelength for each transition is shown by a vertical orange line.}
\label{fig:hrts_spectra}
\end{figure*}

\subsubsection{Results}

\begin{figure*}[!ht]
\centerline{\includegraphics[angle=90,width=9cm,keepaspectratio]{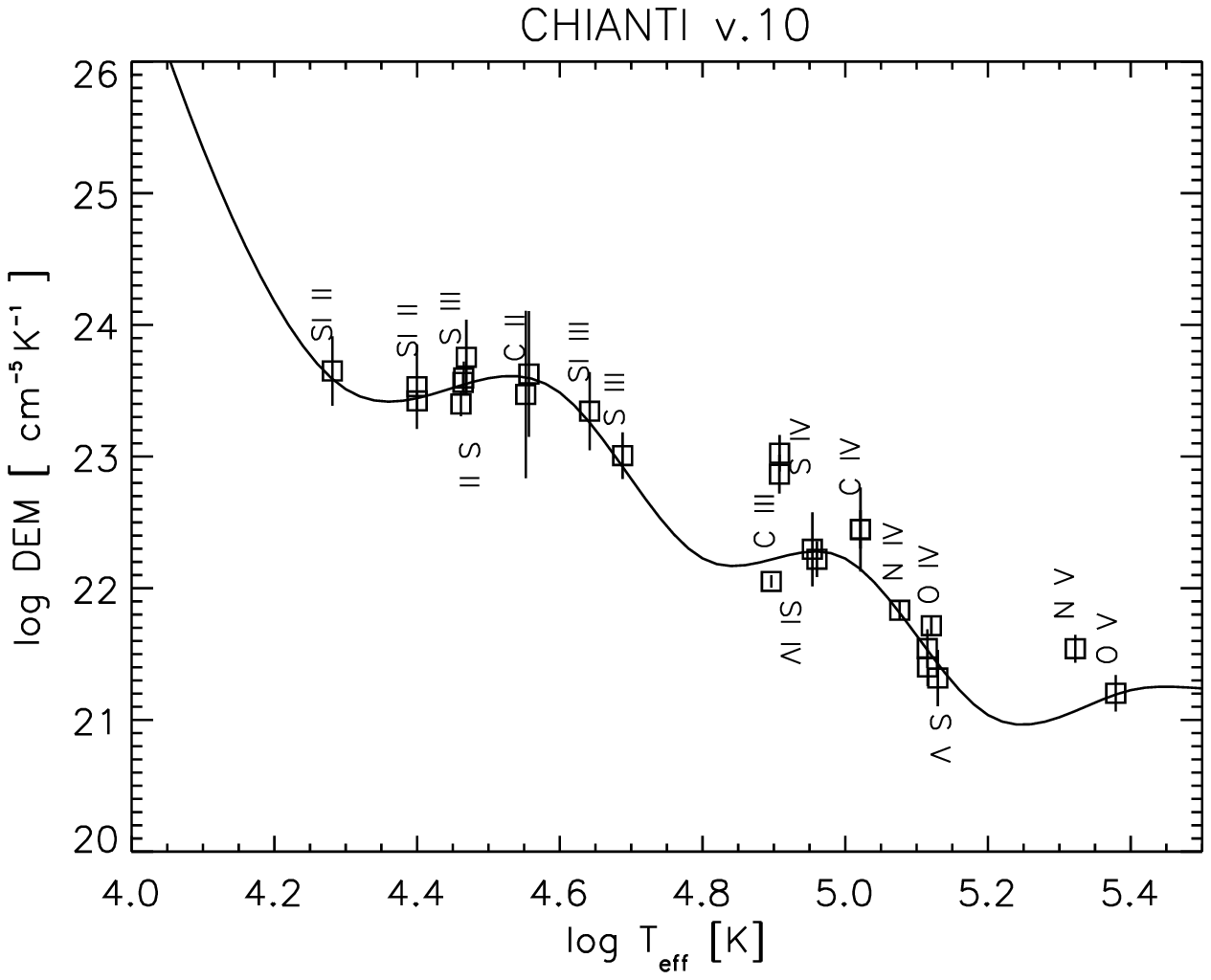}
\includegraphics[angle=90,width=9cm,keepaspectratio]{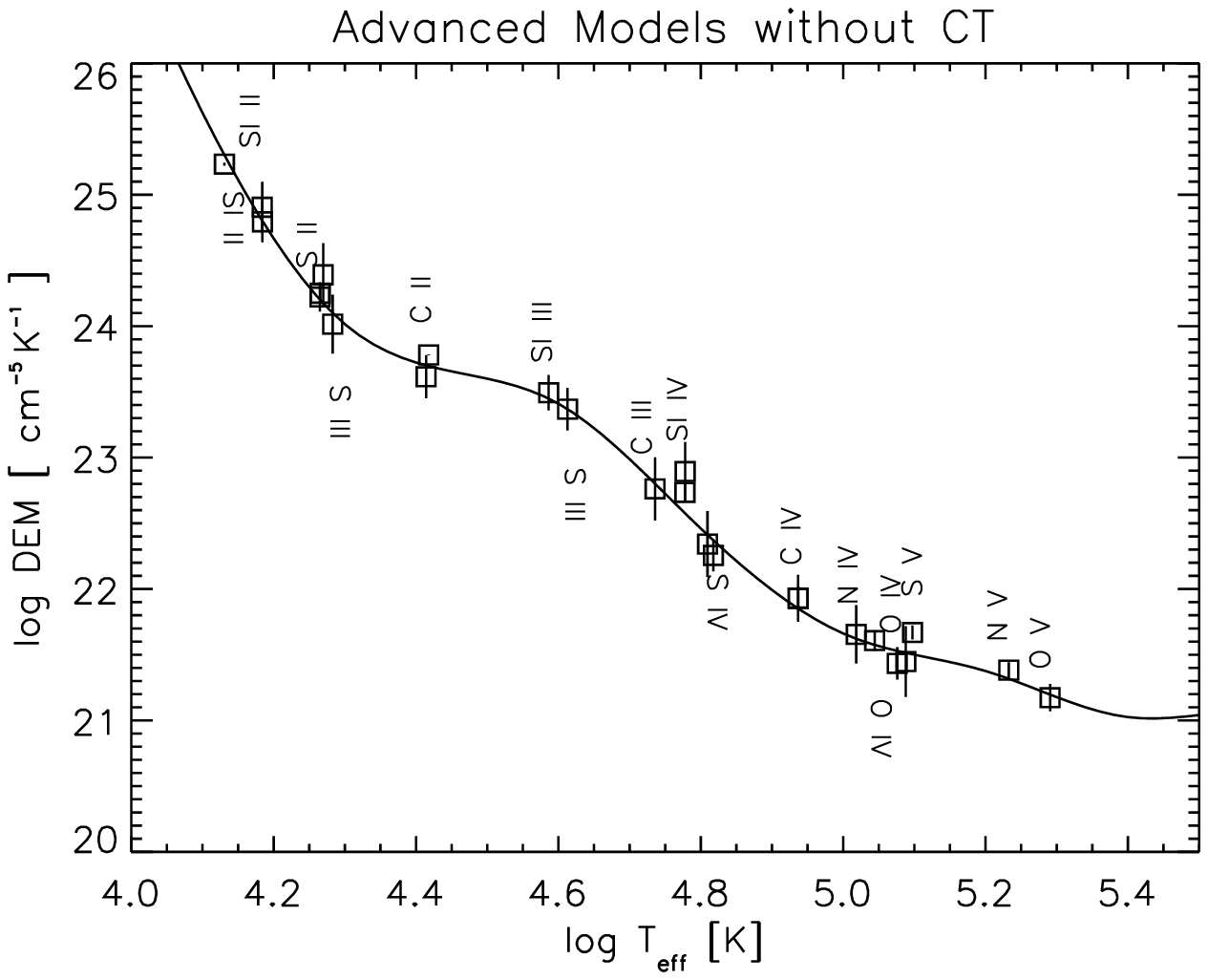}}
\centerline{\includegraphics[angle=90,width=9cm,keepaspectratio]{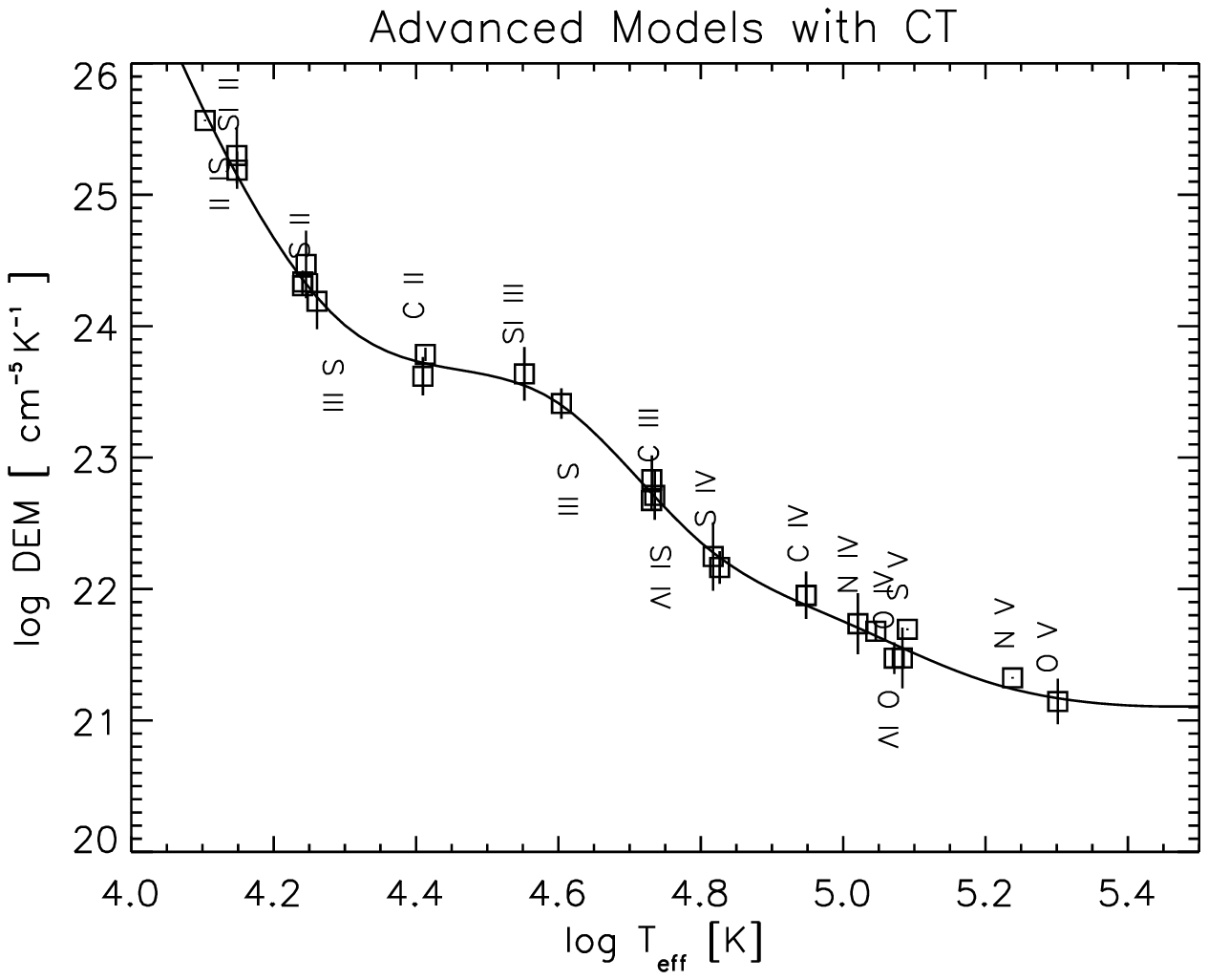}
\includegraphics[angle=0,width=9cm,keepaspectratio]{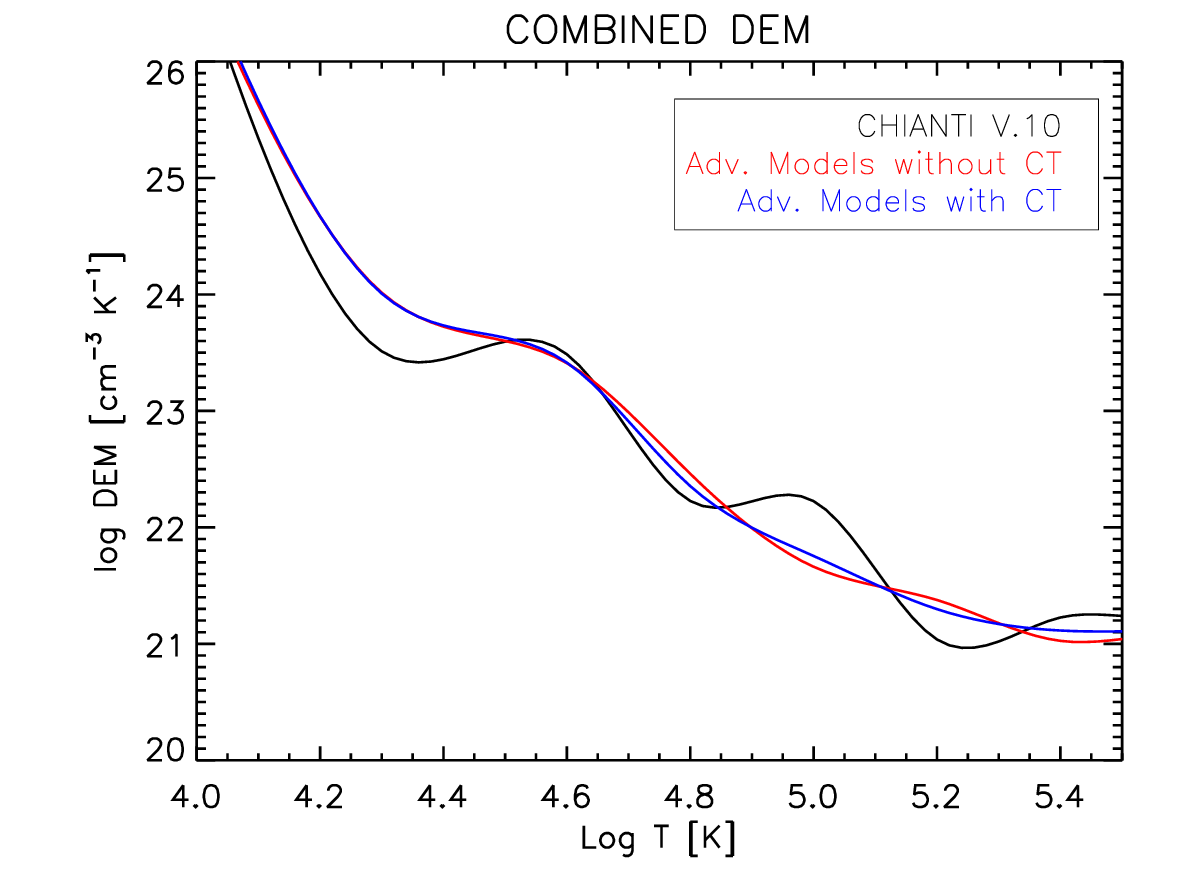}
}
\caption{The DEMs obtained with the three models. The points are plotted at the effective temperature, with values equal to the ratio of observed vs. predicted radiance times the DEM value at the effective temperature.}
\label{fig:dems}
\end{figure*}

Figure~\ref{fig:hrts_spectra} shows four spectral ranges from the HRTS-II spectra with the three model spectra over-plotted. Table~\ref{tab:hrts} shows a summary of the measured radiances, the effective temperatures, and the ratios between predicted and observed values from the three different models. Figure~\ref{fig:dems} shows the resulting DEMs. 

Lines from \ion{Si}{ii} and \ion{S}{ii} are included in the calculation only to constrain the DEM at low temperatures. \citet{lanzafame1994} showed in hydrostatic, radiative transfer calculations that the \ion{Si}{ii} lines are optically thick in all parts of the Sun, and so a DEM calculation will not produce meaningful results for these lines. \citet{dufresne2023piobs} found in the QS case that the \ion{S}{ii} lines were far from the emission measure of other lines forming at a similar temperature. Therefore, neither of these ions will be discussed further other than to say that there is reasonable consistency in the predicted to observed intensities for the lines, which was not the case in the QS observations. 

In high density, optically thin conditions the theoretical ratio of the two \ion{C}{ii} resonance lines should be close to 2. However, it is clear that this is not the case with these observations, for which the ratio is 1.3. While the predicted to observed intensity is unchanged between all three models, it is notable that the line formation temperature decreases from 35\,000\,K in the coronal model to 25\,700\,K in the advanced models. The prediction for the \ion{Si}{iii} 1206.48\,\AA\ resonance line decreases slightly when charge transfer is included in the advance models because the line forms at higher temperature (log\,$T_{\rm eff}=4.55$) than where the ion fraction now peaks in the new models (log\,$T_{\rm eff}=4.45$). There is a slight increase in the predicted intensity of the \ion{S}{iii} line at 1200.94\,\AA\ because of the increase in the peak ion fraction illustrated in Fig.~\ref{fig:ion_balances}.

Clearly, the biggest change in this whole work is with the \ion{Si}{iv} resonance lines. Table~\ref{tab:hrts} and Fig.~\ref{fig:hrts_spectra} show almost a factor of five increase in the intensities of the \ion{Si}{iv} lines in the advanced model with charge transfer, and the outcome is much better agreement with observations. In these observations it appears the 1393.75\,\AA/1402.76\,\AA\ intensity ratio departs from the optically thin value of 2. It suggests opacity is not be the cause of this because opacity would cause the intensity ratio to decrease below 2 as photons are absorbed from the stronger line. \citet{gontikakis2018} analysed IRIS observations and found many areas in an AR where the intensity ratio is greater than 2. The majority of these cases had ratios in the range 2.1-2.6, although they excluded significantly more cases where the signal-to-noise ratio was too low. They attribute line intensity ratios greater than the optically thin limit to resonant scattering of incident radiation in areas of lower electron density.

Figure~\ref{fig:hrts_spectra} shows the much improved ratio of the \ion{S}{iv} lines to the \ion{O}{iv} lines in the IRIS wavelength range. The advanced models appear to resolve the discrepancy in the ratios of these lines noted by \citet{doschek2001}, which limited their analysis of lines which form above and below $10^5$\,K. The present atomic models may eliminate the need for the empirical correction factors used by \citet{young2018iris}. Improvement in this temperature range is also seen in the \ion{C}{iii} line at 1247.39\,\AA. Since carbon is unaffected by CT, the decrease in predicted intensity for this line in the advanced model with CT very likely comes from the effect on the DEM of the \ion{Si}{iv} lines, which form at a similar temperature. 

A notable drop is seen in the formation temperatures of the \ion{S}{iv} lines (from 89\,000\,K to 66\,000\,K); however, the predicted intensities of the \ion{S}{iv}, \ion{N}{iv}, \ion{O}{iv} and \ion{S}{v} lines change by relatively small amounts. This highlights how, generally, differences become smaller between the advanced models and coronal approximation for lines which form closer to the corona, as expected. Although the integrated intensities do not change so much for these ions, the \ion{O}{iv} 1401.16\,\AA\ / \ion{S}{iv} 1406.05\,\AA\ line ratio, which can be used as a temperature diagnostic, was shown by \citet{rao2022} to change by factors of 2-4 when using the advanced models compared to the coronal approximation. 

\begin{table*}[!ht]
\caption{Results of the DEM analysis of the HRTS-II plage. $I_{\text{obs}}$ 
is the measured intensity in erg cm$^{-2}$ s$^{-1}$ sr$^{-1}$.}
\centering
\begin{tabular}{ccccccccc}
\hline
\hline
\multirow{2}{*}{Wavelength (\AA)} & \multirow{2}{*}{$I_{\text{obs}}$} & \multirow{2}{*}{Ion} & \multicolumn{2}{c}{Adv. models with CT} & \multicolumn{2}{c}{Adv. models without CT} & \multicolumn{2}{c}{\chianti\ v.10.1} \\
\cline{4-5} \cline{6-7} \cline{8-9}
& & & $\log T_{\text{eff}}$ & $I_{\text{calc}}/I_{\text{obs}}$ & $\log T_{\text{eff}}$ & $I_{\text{calc}}/I_{\text{obs}}$ & $\log T_{\text{eff}}$ & $I_{\text{calc}}/I_{\text{obs}}$ \\
\hline
1533.44 & 185 & \ion{Si}{ii} & 4.10 & 1.14 & 4.13 & 1.17 & 4.28 & 0.87 \\
1264.74 & 276 & \ion{Si}{ii} & 4.15 & 0.71 & 4.18 & 0.79 & 4.40 & 0.82 \\
1260.42 & 121 & \ion{Si}{ii} & 4.15 & 0.91 & 4.18 & 1.01 & 4.40 & 1.05 \\
1259.52 & 67.2 & \ion{S}{ii} & 4.24 & 1.12 & 4.26 & 0.95 & 4.46 & 0.96 \\
1253.81 & 53.0 & \ion{S}{ii} & 4.24 & 1.04 & 4.27 & 0.89 & 4.47 & 0.89 \\
1250.58 & 37.4 & \ion{S}{ii} & 4.25 & 0.71 & 4.27 & 0.61 & 4.47 & 0.63\\
1194.45 & 134 & \ion{Si}{ii} & 4.26 & 1.07 & 4.28 & 1.22 & 4.46 & 1.38 \\
1335.70 & 3450 & \ion{C}{ii} & 4.41 & 1.26 & 4.41 & 1.22 & 4.55 & 1.35 \\
1334.53 & 2620 & \ion{C}{ii} & 4.41 & 0.85 & 4.42 & 0.82 & 4.56 & 0.93 \\
1206.48 & 4920 & \ion{Si}{iii} & 4.55 & 0.82 & 4.59 & 0.90 & 4.64 & 0.82\\
1200.94 & 110 & \ion{S}{iii} & 4.60 & 0.97 & 4.61 & 1.00 & 4.69 & 0.82 \\
1402.76 & 521 & \ion{Si}{iv} & 4.73 & 1.14 & 4.78 & 0.70 & 4.91 & 0.23 \\
1393.75 & 1500 & \ion{Si}{iv} & 4.73 & 0.79 & 4.78 & 0.48 & 4.91 & 0.16\\
1247.39 & 16.7 & \ion{C}{iii} & 4.74 & 0.98 & 4.74 & 1.09 & 4.90 & 1.47 \\
1423.84 & 5.15 & \ion{S}{iv} & 4.82 & 1.07 & 4.81 & 1.17 & 4.95 & 0.96\\
1406.05 & 26.5 & \ion{S}{iv} & 4.83 & 1.19 & 4.82 & 1.30 & 4.96 & 1.15\\
1550.79 & 1590 & \ion{C}{iv} & 4.95 & 0.84 & 4.94 & 0.83 & 5.02 & 0.50\\
1548.21 & 3170 & \ion{C}{iv} & 4.95 & 0.84 & 4.94 & 0.84 & 5.02 & 0.50 \\
1486.51 & 31.4 & \ion{N}{iv} & 5.02 & 0.93 & 5.02 & 0.93 & 5.08 & 0.96 \\
1404.80 & 36.5 & \ion{O}{iv} & 5.05 & 0.92 & 5.04 & 0.93 & 5.12 & 0.98\\
1407.38 & 26.9 & \ion{O}{iv} & 5.07 & 1.28 & 5.08 & 1.24 & 5.12 & 1.34 \\
1401.16 & 102 & \ion{O}{iv} & 5.08 & 1.19 & 5.09 & 1.17 & 5.13 & 1.28\\
1199.17 & 46.9 & \ion{S}{v} & 5.09 & 0.69 & 5.10 & 0.68 & 5.12 & 0.60\\
1238.82 & 240 & \ion{N}{v} & 5.24 & 0.82 & 5.23 & 0.86 & 5.32 & 0.34 \\
1218.35 & 203 & \ion{O}{v} & 5.30 & 1.06 & 5.29 & 1.06 & 5.38 & 0.98 \\
\hline
\hline
\end{tabular}
\label{tab:hrts}
\end{table*}

There are significant changes in the \ion{C}{iv} and \ion{N}{v} lines, which increase by factors of almost 2 and 3, respectively. They are in much better agreement with observations compared to the coronal approximation. All the anomalous ions are now well represented by the advanced models. The \ion{O}{v} intercombination line at 1218\,\AA\ was a factor of two weaker than observations of the quiet Sun in \citet{dufresne2023piobs}, yet here it is in good agreement with the plage observation. Since no other lines are used to fit the DEM at this and higher temperatures, the DEM may have adjusted to match the observation of this and the \ion{N}{v} line, explaining why they may both be in good agreement with observations for the plage case but were not in the quiet Sun case.

For ions which form at higher temperatures than HRTS was able to observe, such as \ion{Mg}{vii-viii} and \ion{Si}{viii}, the QS line intensities tested by \citet{dufresne2023piobs} were affected by 10-20\% using the advanced models. This also includes the \ion{Ne}{vii} 465.2\,\AA\ line currently being observed by the SUTRI imaging mission \citep{bai2023sutri}; its formation temperature decreases by about 10\% in the advanced models. The only higher temperature lines in the QS which showed greater variations were the Li-like \ion{O}{vi} lines at 1031.9\,\AA\ and 1037.6\,\AA, which were enhanced by just over 40\% using the advanced models. These and the Li-like \ion{Ne}{viii} and Na-like \ion{S}{vi} lines cannot be tested here, but \ion{O}{vi} and \ion{Ne}{viii} are currently observed by SPICE.
As a whole, the models are relevant for a wide wavelength range from past and present missions, including many more lines not covered by the HRTS spectral range used as an illustration here.

\subsubsection{A note on elemental abundances}

What is perhaps surprising when looking at the results as a whole is that good agreement is found using the photospheric elemental abundances of \citet{asplund2021}. It is often assumed in the literature that the abundances of elements with low first ionization potential (FIP), such as silicon, are greater in the atmosphere relative to the abundances of high-FIP elements, such as C, N and O. In active regions typical enhancements are assumed for low FIP elements by factors between 2 and 4 relative to photospheric abundance ratios \citep[see the reviews by][]{laming2015,delzanna2018}. This is the so-called FIP effect. It is quite well established in quiescent ARs that high temperature plasma around 3$\times10^6$\,K plasma has a FIP effect; the 1$\times10^6$\,K plasma in ARs has instead shown a wide range of results.

\cite{feldman_etal:1990} used the HRTS-I observations to argue that the FIP effect is also present in the chromosphere/low transition region in an active region, where they mostly looked at Si/C intensity ratios. However, these were simply comments based on the appearance of the monochromatic images, assuming that the ions were formed at the temperatures calculated using the coronal approximation. The present work has shown clearly that assuming such temperatures is incorrect. \cite{doschek_etal:1991} also analysed this observation and found the same variations in the Si/C intensity ratios, but were more cautious in their conclusions. 

For the specific HRTS-II plage observation used here, if the relative Si/C abundance was increased by a factor of 2--4, the \ion{Si}{iii} and \ion{Si}{iv} lines would become over-predicted by the same amount. Sulphur has an FIP of 10\,eV but in remote-sensing observations has an abundance which follows that of the high-FIP elements, depending on the conditions which lead to fractionation and the FIP effect \citep{laming2019}. The sulphur lines in this work are generally weak but are well represented by photospheric abundances. A wider selection of lines for the TR than are available with HRTS may be required, but clearly this issue may warrant further investigation using advanced atomic models.

\section{New and updated ions}
\label{sec:otherdata}

\subsection{{\rm \ion{Ne}{i}}}

Details about this new addition are given in Sect.~\ref{sec:ci}.

\subsection{{\rm \ion{Mg}{vii}}}

\citet{2023ApJ...958...40Y} provided updated experimental energies for the $2s^22p^2$ $^3P_{1,2}$ and $^1D_2$ levels (indices 2--4), and the $2s2p^3$ $^3S_1$ and $^1P_1$ levels (indices 14, 15) and these have been added to the CHIANTI energy level file. The wavelengths in the radiative data file have been recomputed from the new energies.

\subsection{{\rm \ion{Si}{vii}}}

\citet{2023ApJ...958...40Y} provided updated experimental energies for the 
$2s^22p^4$ $^3P_{1,0}$ levels  (indices 2,3), and the $2s2p^5$ $^3P_{2,1,0}$  levels (indices 6--8) and these have been added to the CHIANTI energy level file. The wavelengths in the radiative data file have been recomputed from the new energies.

\subsection{{\rm \ion{Ar}{v}} and {\rm \ion{Ar}{vi}}}

The ions \ion{Ar}{v} and \ion{Ar}{vi} are of interest because they have been observed in SUMER spectra.  They are examples of high FIP ions and can be used to study variations of elemental abundances.  The previous version of \chianti\ did not contain the ion \ion{Ar}{vi} and the ion \ion{Ar}{v} only included five levels in the ground configuration.

Theoretical energy levels, A-values, and effective collision strengths have been calculated for the Argon iso-nuclear sequence by \citet{ludlow_ar} and these have provided the atomic parameters for creating the new and updated models.  These parameters have been supplemented by theoretical energy levels from \citet{nist}.  Wavelengths are derived from the energy levels and include ``observed" wavelengths when there are NIST values for the upper and lower levels.  Otherwise, the wavelengths are derived from the theoretical energies and are considered to be ``unobserved."  The high temperature limit for the collision strengths are calculated from the weighted oscillator strength (\textit{gf}-value) following \citet{burly} with the electric dipole values provided by an \textsc{Autostructure} calculation \citep{badnell2011}.

\subsection{{\rm \ion{Fe}{xiv}} and {\rm \ion{Fe}{xv}}}

Recently \citet{lepson_fe} have reported laboratory wavelengths of EUV lines of \ion{Fe}{xiv--xvi}.  These have allowed some new values for energies and wavelengths for \ion{Fe}{xiv} and \ion{Fe}{xv} but not for \ion{Fe}{xvi}.  In the case of \ion{Fe}{xiv}, \citet{lepson_fe} identify a line at 55.152 \AA\ that we listed as the transition 3s$^2$5f $^2$F$_{5/2}$ to 3s$^2$3d $^2$D$_{3/2}$ and provide an ``observed" energy for the upper level.  For \ion{Fe}{xv}, they report a line at 53.10 \AA\ that is identified as the transition between the levels 3s$^2$ $^1$S$_{0}$ and $3s4p$ $^3P_{1}$ and a line at 46.14 \AA\ that is identified as a transition between the 3s3p $^1$P$_{1}$ and 3s5s $^1$S$_0$ levels.  These allow us to assign  ``observed" energies to the upper levels.  In addition, the NIST energies for the levels 3s4f $^3$F$_2$  and 3s4f $^3$F$_3$ have been re-inserted and the value for the 3p4f $^3$G$_5$ is a new assignment.

\section{Improved radiative recombination energy loss rates}

\citet{mao_rrloss} have used the photoionization cross sections of \citet{badnell2006} to derive the radiative recombination energy loss rate for ions in the H-like through Ne-like isoelectronic sequence with Z between 1 and 30. The authors provide tables of parameterized fits to the energy loss rates.  These fit parameters are now included in the \chianti\ database for the appropriate ions.  The greatest changes are at low temperatures where the rates of \citet{mao_rrloss} are considerably larger than our previous values, which relied on the \citet{karzas1961} free-bound Gaunt factors for levels with principal quantum numbers up to six.  The \citet{mao_rrloss} rates include recombination to higher levels that are important at lower temperatures.

\section{Conclusions}  \label{sec:conclusions}

The present advanced models provide a significant change in the formation temperature of all the ions in the transition region, and cause a large increase in the predicted intensities of the anomalous ions. The physics of these effects have been known for a long time, but previous models adopted more simplified atomic rates than are provided here. The present models do have different levels of approximations, but are sufficient to clearly indicate where the added physical effects are most important. 

An example observation is provided here to illustrate the improved agreement in the TR lines. The earlier comparison with quiet Sun observations in \citet{dufresne2023piobs}, using the same models but with a wider selection of lines, found cases where significant discrepancies in the anomalous ions were still present. Comparisons with observations prior to that have also indicated an unclear picture, with the simple, static-atmosphere DEM models able in some cases to resolve the long-standing discrepancies for the anomalous ions. There are more factors to consider when modelling spectral lines in the transition region, such as time dependent ionization, radiative transfer and non-Maxwellian electron distributions. This should caution the reader that other effects are present.

The present advanced models are provided as a useful tool for any physics-based models of the solar transition region, and also as a tool for interpreting emission from higher density astrophysical and laboratory plasma. For example, \citet{metcalfe2023} carried out density diagnostics of the transition region of an exoplanet host star and found similar conditions to the Sun. In their analysis, they use many of the lines for which the present models are intended, and which have shown significant variations compared to the coronal approximation.

We note that photo-ionization and photo-excitation can also be important effects. This affects many TR lines, altering in some cases emission in the quiet Sun from singly- and doubly-charged ions by factors of 2-7, such as the intercombination lines from \ion{C}{ii} and \ion{O}{iii} \citep{dufresne2023piobs}. These processes are also important in the outer, low-density corona. Advanced models including such processes will be included in a future \chianti\ release.

\begin{acknowledgments}
GDZ and RPD acknowledge support from STFC (UK) via the consolidated grants to the atomic astrophysics group at DAMTP, University of Cambridge (ST/P000665/1. and  ST/T000481/1). PRY acknowledges support from the NASA Heliophysics Digital Resource Library. 
ED acknowledges support from STFC (UK) via a studentship.  KPD acknowledges support from NASA grants 80NSSC21K1785 and 80NSSC24K0119.  
We thank Professor Klaus Bartschat and Dr Yang Wang for providing excitation and ionization data for neutrals from their published works.
The UK APAP network (PI: N. Badnell), also funded over the years by STFC,  has provided a large proportion of the atomic data currently present in the database and used for the advanced models.
\end{acknowledgments}

\bibliography{paper}{}

\begin{thebibliography}{}
\expandafter\ifx\csname natexlab\endcsname\relax\def\natexlab#1{#1}\fi
\providecommand{\url}[1]{\href{#1}{#1}}
\providecommand{\dodoi}[1]{doi:~\href{http://doi.org/#1}{\nolinkurl{#1}}}
\providecommand{\doeprint}[1]{\href{http://ascl.net/#1}{\nolinkurl{http://ascl.net/#1}}}
\providecommand{\doarXiv}[1]{\href{https://arxiv.org/abs/#1}{\nolinkurl{https://arxiv.org/abs/#1}}}

\bibitem[{{Asplund} {et~al.}(2021){Asplund}, {Amarsi}, \& {Grevesse}}]{asplund2021}
{Asplund}, M., {Amarsi}, A.~M., \& {Grevesse}, N. 2021, \aap, 653, A141, \dodoi{10.1051/0004-6361/202140445}

\bibitem[{{Avrett} \& {Loeser}(2008)}]{avrett2008}
{Avrett}, E.~H., \& {Loeser}, R. 2008, \apjs, 175, 229, \dodoi{10.1086/523671}

\bibitem[{{Bacchus-Montabonel} \& {Amezian}(1993)}]{bacchus1993s3}
{Bacchus-Montabonel}, M.~C., \& {Amezian}, K. 1993, Zeitschrift fur Physik D Atoms Molecules Clusters, 25, 323, \dodoi{10.1007/BF01437298}

\bibitem[{{Badnell}(2006)}]{badnell2006}
{Badnell}, N.~R. 2006, \apjs, 167, 334, \dodoi{10.1086/508465}

\bibitem[{{Badnell}(2011)}]{badnell2011}
---. 2011, Computer Physics Communications, 182, 1528, \dodoi{10.1016/j.cpc.2011.03.023}

\bibitem[{{Badnell} {et~al.}(2003){Badnell}, {O'Mullane}, {Summers}, {Altun}, {Bautista}, {Colgan}, {Gorczyca}, {Mitnik}, {Pindzola}, \& {Zatsarinny}}]{badnell2003}
{Badnell}, N.~R., {O'Mullane}, M.~G., {Summers}, H.~P., {et~al.} 2003, \aap, 406, 1151, \dodoi{10.1051/0004-6361:20030816}

\bibitem[{{Bai} {et~al.}(2023){Bai}, {Tian}, {Deng}, {Wang}, {Yang}, {Zhang}, {Zhang}, {Qi}, {Wang}, {Gao}, {Yu}, {He}, {Shen}, {Shen}, {Guo}, {Hou}, {Ji}, {Bi}, {Duan}, {Yang}, {Lin}, {Hu}, {Song}, {Yang}, {Chen}, {Qiao}, {Ge}, {Li}, {Jin}, {He}, {Chen}, {Zhu}, {He}, {Shi}, {Liu}, {Li}, {Xu}, {Liu}, {Li}, {Feng}, {Wang}, {Fan}, {Liu}, {Guo}, {Sun}, {Wu}, {Li}, {Yang}, {Ye}, {Gu}, {Wu}, {Zhang}, {Yu}, {Ye}, {Sheng}, {Wang}, {Li}, {Huang}, \& {Zhang}}]{bai2023sutri}
{Bai}, X., {Tian}, H., {Deng}, Y., {et~al.} 2023, Research in Astronomy and Astrophysics, 23, 065014, \dodoi{10.1088/1674-4527/accc74}

\bibitem[{{Baliunas} \& {Butler}(1980)}]{baliunas1980}
{Baliunas}, S.~L., \& {Butler}, S.~E. 1980, \apjl, 235, L45, \dodoi{10.1086/183154}

\bibitem[{{Barklem} {et~al.}(2017){Barklem}, {Osorio}, {Fursa}, {Bray}, {Zatsarinny}, {Bartschat}, \& {Jerkstrand}}]{barklem2017}
{Barklem}, P.~S., {Osorio}, Y., {Fursa}, D.~V., {et~al.} 2017, \aap, 606, A11, \dodoi{10.1051/0004-6361/201730864}

\bibitem[{{Barrag{\'a}n} {et~al.}(2006){Barrag{\'a}n}, {Errea}, {M{\'e}ndez}, {Rabad{\'a}n}, \& {Riera}}]{barragan2006}
{Barrag{\'a}n}, P., {Errea}, L.~F., {M{\'e}ndez}, L., {Rabad{\'a}n}, I., \& {Riera}, A. 2006, \apj, 636, 544, \dodoi{10.1086/497884}

\bibitem[{{Bates} \& {McCarroll}(1962)}]{bates1962ct}
{Bates}, D.~R., \& {McCarroll}, R. 1962, Advances in Physics, 11, 39, \dodoi{10.1080/00018736200101262}

\bibitem[{{Bienstock} {et~al.}(1984){Bienstock}, {Dalgarno}, \& {Heil}}]{bienstock1984}
{Bienstock}, S., {Dalgarno}, A., \& {Heil}, T.~G. 1984, \pra, 29, 2239, \dodoi{10.1103/PhysRevA.29.2239}

\bibitem[{{Brekke}(1993)}]{brekke1993}
{Brekke}, P. 1993, \apjs, 87, 443, \dodoi{10.1086/191810}

\bibitem[{{Burgess} \& {Chidichimo}(1983)}]{burgess1983}
{Burgess}, A., \& {Chidichimo}, M.~C. 1983, \mnras, 203, 1269, \dodoi{10.1093/mnras/203.4.1269}

\bibitem[{{Burgess} \& {Summers}(1969)}]{burgess1969}
{Burgess}, A., \& {Summers}, H.~P. 1969, \apj, 157, 1007, \dodoi{10.1086/150131}

\bibitem[{{Burgess} \& {Summers}(1976)}]{burgess1976}
---. 1976, \mnras, 174, 345, \dodoi{10.1093/mnras/174.2.345}

\bibitem[{{Burgess} \& {Tully}(1992)}]{burly}
{Burgess}, A., \& {Tully}, J.~A. 1992, \aap, 254, 436

\bibitem[{Burton {et~al.}(1971)Burton, Jordan, Ridgeley, \& Wilson}]{burton1971}
Burton, W.~M., Jordan, C., Ridgeley, A., \& Wilson, R. 1971, Philos. Trans. R. Soc. London, Ser. A, 270, 81

\bibitem[{{Butler} \& {Dalgarno}(1980)}]{butler1980}
{Butler}, S.~E., \& {Dalgarno}, A. 1980, \apj, 241, 838, \dodoi{10.1086/158395}

\bibitem[{{Chatzikos} {et~al.}(2023){Chatzikos}, {Bianchi}, {Camilloni}, {Chakraborty}, {Gunasekera}, {Guzm{\'a}n}, {Milby}, {Sarkar}, {Shaw}, {van Hoof}, \& {Ferland}}]{chatzikos2023}
{Chatzikos}, M., {Bianchi}, S., {Camilloni}, F., {et~al.} 2023, \rmxaa, 59, 327, \dodoi{10.22201/ia.01851101p.2023.59.02.12}

\bibitem[{{Christensen} \& {Watson}(1981)}]{christensen1981}
{Christensen}, R.~B., \& {Watson}, W.~D. 1981, \pra, 24, 1331, \dodoi{10.1103/PhysRevA.24.1331}

\bibitem[{{Clarke} {et~al.}(1998){Clarke}, {Stancil}, {Zygelman}, \& {Cooper}}]{clarke1998}
{Clarke}, N.~J., {Stancil}, P.~C., {Zygelman}, B., \& {Cooper}, D.~L. 1998, Journal of Physics B Atomic Molecular Physics, 31, 533, \dodoi{10.1088/0953-4075/31/3/019}

\bibitem[{De~Pontieu {et~al.}(2014)De~Pontieu, Title, Lemen, Kushner, Akin, Allard, Berger, Boerner, Cheung, Chou, Drake, Duncan, Freeland, Heyman, Hoffman, Hurlburt, Lindgren, Mathur, Rehse, Sabolish, Seguin, Schrijver, Tarbell, W{\"u}lser, Wolfson, Yanari, Mudge, Nguyen-Phuc, Timmons, van Bezooijen, Weingrod, Brookner, Butcher, Dougherty, Eder, Knagenhjelm, Larsen, Mansir, Phan, Boyle, Cheimets, DeLuca, Golub, Gates, Hertz, McKillop, Park, Perry, Podgorski, Reeves, Saar, Testa, Tian, Weber, Dunn, Eccles, Jaeggli, Kankelborg, Mashburn, Pust, Springer, Carvalho, Kleint, Marmie, Mazmanian, Pereira, Sawyer, Strong, Worden, Carlsson, Hansteen, Leenaarts, Wiesmann, Aloise, Chu, Bush, Scherrer, Brekke, Martinez-Sykora, Lites, McIntosh, Uitenbroek, Okamoto, Gummin, Auker, Jerram, Pool, \& Waltham}]{depontieu2014}
De~Pontieu, B., Title, A.~M., Lemen, J.~R., {et~al.} 2014, \solphys, 289, 2733, \dodoi{10.1007/s11207-014-0485-y}

\bibitem[{{Del Zanna} {et~al.}(2021){Del Zanna}, {Dere}, {Young}, \& {Landi}}]{delzanna2021v10}
{Del Zanna}, G., {Dere}, K.~P., {Young}, P.~R., \& {Landi}, E. 2021, \apj, 909, 38, \dodoi{10.3847/1538-4357/abd8ce}

\bibitem[{Del~Zanna {et~al.}(2002)Del~Zanna, Landini, \& Mason}]{delzanna2002}
Del~Zanna, G., Landini, M., \& Mason, H.~E. 2002, \aap, 385, 968, \dodoi{10.1051/0004-6361:20020164}

\bibitem[{{Del Zanna} \& {Mason}(2018)}]{delzanna2018}
{Del Zanna}, G., \& {Mason}, H.~E. 2018, Living Reviews in Solar Physics, 15, 5, \dodoi{10.1007/s41116-018-0015-3}

\bibitem[{{Del Zanna} {et~al.}(2020){Del Zanna}, {Storey}, {Badnell}, \& {Andretta}}]{delzanna2020}
{Del Zanna}, G., {Storey}, P.~J., {Badnell}, N.~R., \& {Andretta}, V. 2020, \apj, 898, 72, \dodoi{10.3847/1538-4357/ab9d84}

\bibitem[{{Dere}(2007)}]{dere2007}
{Dere}, K.~P. 2007, \aap, 466, 771, \dodoi{10.1051/0004-6361:20066728}

\bibitem[{{Dere} {et~al.}(2023){Dere}, {Del Zanna}, {Young}, \& {Landi}}]{dere2023}
{Dere}, K.~P., {Del Zanna}, G., {Young}, P.~R., \& {Landi}, E. 2023, \apjs, 268, 52, \dodoi{10.3847/1538-4365/acec79}

\bibitem[{{Dere} {et~al.}(1997){Dere}, {Landi}, {Mason}, {Monsignori Fossi}, \& {Young}}]{dere1997}
{Dere}, K.~P., {Landi}, E., {Mason}, H.~E., {Monsignori Fossi}, B.~C., \& {Young}, P.~R. 1997, \aaps, 125, 149, \dodoi{10.1051/aas:1997368}

\bibitem[{{Doschek} {et~al.}(1991){Doschek}, {Dere}, \& {Lund}}]{doschek_etal:1991}
{Doschek}, G.~A., {Dere}, K.~P., \& {Lund}, P.~A. 1991, \apj, 381, 583, \dodoi{10.1086/170683}

\bibitem[{{Doschek} \& {Mariska}(2001)}]{doschek2001}
{Doschek}, G.~A., \& {Mariska}, J.~T. 2001, \apj, 560, 420, \dodoi{10.1086/322771}

\bibitem[{Doyle {et~al.}(2005)Doyle, Summers, \& Bryans}]{doyle2005}
Doyle, J.~G., Summers, H.~P., \& Bryans, P. 2005, \aap, 430, L29, \dodoi{10.1051/0004-6361:200400125}

\bibitem[{{Dud{\'{\i}}k} {et~al.}(2014){Dud{\'{\i}}k}, {Del Zanna}, {Dzif{\v c}{\'a}kov{\'a}}, {Mason}, \& {Golub}}]{dudik_etal:2014_o_4}
{Dud{\'{\i}}k}, J., {Del Zanna}, G., {Dzif{\v c}{\'a}kov{\'a}}, E., {Mason}, H.~E., \& {Golub}, L. 2014, \apjl, 780, L12, \dodoi{10.1088/2041-8205/780/1/L12}

\bibitem[{{Dufresne} \& {Del Zanna}(2019)}]{dufresne2019}
{Dufresne}, R.~P., \& {Del Zanna}, G. 2019, \aap, 626, A123, \dodoi{10.1051/0004-6361/201935133}

\bibitem[{{Dufresne} {et~al.}(2020){Dufresne}, {Del Zanna}, \& {Badnell}}]{dufresne2020}
{Dufresne}, R.~P., {Del Zanna}, G., \& {Badnell}, N.~R. 2020, \mnras, 497, 1443, \dodoi{10.1093/mnras/staa2005}

\bibitem[{{Dufresne} {et~al.}(2021{\natexlab{a}}){Dufresne}, {Del Zanna}, \& {Badnell}}]{dufresne2021pico}
---. 2021{\natexlab{a}}, \mnras, 503, 1976, \dodoi{10.1093/mnras/stab514}

\bibitem[{{Dufresne} {et~al.}(2023){Dufresne}, {Del Zanna}, \& {Mason}}]{dufresne2023piobs}
{Dufresne}, R.~P., {Del Zanna}, G., \& {Mason}, H.~E. 2023, \mnras, 521, 4696, \dodoi{10.1093/mnras/stad794}

\bibitem[{{Dufresne} {et~al.}(2021{\natexlab{b}}){Dufresne}, {Del Zanna}, \& {Storey}}]{dufresne2021picrm}
{Dufresne}, R.~P., {Del Zanna}, G., \& {Storey}, P.~J. 2021{\natexlab{b}}, \mnras, 505, 3968, \dodoi{10.1093/mnras/stab1498}

\bibitem[{{Dupree}(1972)}]{dupree1972}
{Dupree}, A.~K. 1972, \apj, 178, 527, \dodoi{10.1086/151813}

\bibitem[{{Errea} {et~al.}(2015){Errea}, {Illescas}, {Jorge}, {M{\'e}ndez}, {Rabad{\'a}n}, \& {Su{\'a}rez}}]{errea2015}
{Errea}, L.~F., {Illescas}, C., {Jorge}, A., {et~al.} 2015, in Journal of Physics Conference Series, Vol. 576, Journal of Physics Conference Series, 012002, \dodoi{10.1088/1742-6596/576/1/012002}

\bibitem[{{Errea} {et~al.}(2000){Errea}, {Mac{\'\i}as}, {M{\'e}ndez}, \& {Riera}}]{errea2000}
{Errea}, L.~F., {Mac{\'\i}as}, A., {M{\'e}ndez}, L., \& {Riera}, A. 2000, Journal of Physics B Atomic Molecular Physics, 33, 1369, \dodoi{10.1088/0953-4075/33/7/306}

\bibitem[{{Favreau} {et~al.}(2019){Favreau}, {Johnson}, {Ennis}, \& {Loch}}]{favreau2019}
{Favreau}, C.~J., {Johnson}, C.~A., {Ennis}, D.~A., \& {Loch}, S.~D. 2019, Journal of Physics B Atomic Molecular Physics, 52, 095203, \dodoi{10.1088/1361-6455/ab13f4}

\bibitem[{{Feldman} {et~al.}(1990){Feldman}, {Widing}, \& {Lund}}]{feldman_etal:1990}
{Feldman}, U., {Widing}, K.~G., \& {Lund}, P.~A. 1990, \apjl, 364, L21, \dodoi{10.1086/185865}

\bibitem[{{Fischer}(2005)}]{fischer2005}
{Fischer}, C.~F. 2005, \pra, 71, 042506, \dodoi{10.1103/PhysRevA.71.042506}

\bibitem[{{Fontenla} {et~al.}(2014){Fontenla}, {Landi}, {Snow}, \& {Woods}}]{fontenla2014}
{Fontenla}, J.~M., {Landi}, E., {Snow}, M., \& {Woods}, T. 2014, \solphys, 289, 515, \dodoi{10.1007/s11207-013-0431-4}

\bibitem[{{Froese Fischer} \& {Tachiev}(2004)}]{froese2004}
{Froese Fischer}, C., \& {Tachiev}, G. 2004, Atomic Data and Nuclear Data Tables, 87, 1, \dodoi{10.1016/j.adt.2004.02.001}

\bibitem[{{Froese Fischer} {et~al.}(2007){Froese Fischer}, {Tachiev}, {Gaigalas}, \& {Godefroid}}]{fischer2007atsp}
{Froese Fischer}, C., {Tachiev}, G., {Gaigalas}, G., \& {Godefroid}, M.~R. 2007, Computer Physics Communications, 176, 559, \dodoi{10.1016/j.cpc.2007.01.006}

\bibitem[{{Gargaud} {et~al.}(1981){Gargaud}, {Hanssen}, {McCarroll}, \& {Valiron}}]{gargaud1981}
{Gargaud}, M., {Hanssen}, J., {McCarroll}, R., \& {Valiron}, P. 1981, Journal of Physics B Atomic Molecular Physics, 14, 2259, \dodoi{10.1088/0022-3700/14/13/022}

\bibitem[{{Gontikakis} \& {Vial}(2018)}]{gontikakis2018}
{Gontikakis}, C., \& {Vial}, J.~C. 2018, \aap, 619, A64, \dodoi{10.1051/0004-6361/201732563}

\bibitem[{{Gu}(2008)}]{gu2008}
{Gu}, M.~F. 2008, Canadian Journal of Physics, 86, 675, \dodoi{10.1139/P07-197}

\bibitem[{{Imai} {et~al.}(2003){Imai}, {Kimura}, {Gu}, {Hirsch}, {Buenker}, {Wang}, {Stancil}, \& {Pichl}}]{imai2003}
{Imai}, T., {Kimura}, M., {Gu}, J., {et~al.} 2003, \pra, 68, 012716, \dodoi{10.1103/PhysRevA.68.012716}

\bibitem[{{Janev} {et~al.}(1988){Janev}, {Phaneuf}, \& {Hunter}}]{janev1988}
{Janev}, R.~K., {Phaneuf}, R.~A., \& {Hunter}, H.~T. 1988, Atomic Data and Nuclear Data Tables, 40, 249, \dodoi{10.1016/0092-640X(88)90008-3}

\bibitem[{{Jordan}(1969)}]{jordan1969}
{Jordan}, C. 1969, \mnras, 142, 501, \dodoi{10.1093/mnras/142.4.501}

\bibitem[{{Judge} {et~al.}(1995){Judge}, {Woods}, {Brekke}, \& {Rottman}}]{judge1995}
{Judge}, P.~G., {Woods}, T.~N., {Brekke}, P., \& {Rottman}, G.~J. 1995, \apjl, 455, L85, \dodoi{10.1086/309815}

\bibitem[{{Kambara} {et~al.}(2021){Kambara}, {Kawate}, {Oishi}, {Kawamoto}, {Sakaue}, {Kato}, {Nakamura}, {Hara}, \& {Murakami}}]{kambara2021}
{Kambara}, N., {Kawate}, T., {Oishi}, T., {et~al.} 2021, Atoms, 9, 60, \dodoi{10.3390/atoms9030060}

\bibitem[{{Karzas} \& {Latter}(1961)}]{karzas1961}
{Karzas}, W.~J., \& {Latter}, R. 1961, \apjs, 6, 167, \dodoi{10.1086/190063}

\bibitem[{{Kimura} {et~al.}(1997){Kimura}, {Gu}, {Hirsch}, \& {Buenker}}]{kimura1997}
{Kimura}, M., {Gu}, J.~P., {Hirsch}, G., \& {Buenker}, R.~J. 1997, \pra, 55, 2778, \dodoi{10.1103/PhysRevA.55.2778}

\bibitem[{{Kimura} {et~al.}(1996){Kimura}, {Sannigrahi}, {Gu}, {Hirsch}, {Buenker}, \& {Shimamura}}]{kimura1996si1}
{Kimura}, M., {Sannigrahi}, A.~B., {Gu}, J.~P., {et~al.} 1996, \apj, 473, 1114, \dodoi{10.1086/178221}

\bibitem[{{Kingdon} \& {Ferland}(1996)}]{kingdon1996}
{Kingdon}, J.~B., \& {Ferland}, G.~J. 1996, \apjs, 106, 205, \dodoi{10.1086/192335}

\bibitem[{Kramida {et~al.}(2023)Kramida, {Yu.~Ralchenko}, Reader, \& {and NIST ASD Team}}]{nist}
Kramida, A., {Yu.~Ralchenko}, Reader, J., \& {and NIST ASD Team}. 2023, {NIST Atomic Spectra Database (ver. 5.11), [Online]. Available: {\tt{https://physics.nist.gov/asd}} [2024, January 4]. National Institute of Standards and Technology, Gaithersburg, MD.}

\bibitem[{{Laming}(2015)}]{laming2015}
{Laming}, J.~M. 2015, Living Reviews in Solar Physics, 12, 2, \dodoi{10.1007/lrsp-2015-2}

\bibitem[{{Laming} {et~al.}(2019){Laming}, {Vourlidas}, {Korendyke}, {Chua}, {Cranmer}, {Ko}, {Kuroda}, {Provornikova}, {Raymond}, {Raouafi}, {Strachan}, {Tun-Beltran}, {Weberg}, \& {Wood}}]{laming2019}
{Laming}, J.~M., {Vourlidas}, A., {Korendyke}, C., {et~al.} 2019, \apj, 879, 124, \dodoi{10.3847/1538-4357/ab23f1}

\bibitem[{{Lanzafame}(1994)}]{lanzafame1994}
{Lanzafame}, A.~C. 1994, \aap, 287, 972

\bibitem[{{Lepson} {et~al.}(2023){Lepson}, {Beiersdorfer}, {Brown}, \& {Liedahl}}]{lepson_fe}
{Lepson}, J.~K., {Beiersdorfer}, P., {Brown}, G.~V., \& {Liedahl}, D.~A. 2023, \apj, 946, 23, \dodoi{10.3847/1538-4357/acbc17}

\bibitem[{{Lin} {et~al.}(2005){Lin}, {Stancil}, {Gu}, {Buenker}, \& {Kimura}}]{lin2005}
{Lin}, C.~Y., {Stancil}, P.~C., {Gu}, J.~P., {Buenker}, R.~J., \& {Kimura}, M. 2005, \pra, 71, 062708, \dodoi{10.1103/PhysRevA.71.062708}

\bibitem[{{Liu} {et~al.}(2003){Liu}, {Le}, \& {Lin}}]{liu2003}
{Liu}, C.-N., {Le}, A.-T., \& {Lin}, C.~D. 2003, \pra, 68, 062702, \dodoi{10.1103/PhysRevA.68.062702}

\bibitem[{{Liu} {et~al.}(2010{\natexlab{a}}){Liu}, {Qu}, {Xiao}, {Liu}, {Zhou}, {Wang}, \& {Buenker}}]{liu2010ne1rad}
{Liu}, X.~J., {Qu}, Y.~Z., {Xiao}, B.~J., {et~al.} 2010{\natexlab{a}}, \pra, 81, 022717, \dodoi{10.1103/PhysRevA.81.022717}

\bibitem[{{Liu} {et~al.}(2010{\natexlab{b}}){Liu}, {Qu}, {Xiao}, {Liu}, {Zhou}, {Wang}, \& {Buenker}}]{liu2010ne1col}
---. 2010{\natexlab{b}}, Journal of Physics B Atomic Molecular Physics, 43, 085207, \dodoi{10.1088/0953-4075/43/8/085207}

\bibitem[{{Liu} {et~al.}(2011){Liu}, {Wang}, {Qu}, \& {Buenker}}]{liu2011}
{Liu}, X.~J., {Wang}, J.~G., {Qu}, Y.~Z., \& {Buenker}, R.~J. 2011, \pra, 84, 042706, \dodoi{10.1103/PhysRevA.84.042706}

\bibitem[{{Ludlow} {et~al.}(2010){Ludlow}, {Ballance}, {Loch}, \& {Pindzola}}]{ludlow_ar}
{Ludlow}, J.~A., {Ballance}, C.~P., {Loch}, S.~D., \& {Pindzola}, M.~S. 2010, Journal of Physics B Atomic Molecular Physics, 43, 074029, \dodoi{10.1088/0953-4075/43/7/074029}

\bibitem[{{Ludlow} {et~al.}(2008){Ludlow}, {Loch}, {Pindzola}, {Ballance}, {Griffin}, {Bannister}, \& {Fogle}}]{ludlow2008}
{Ludlow}, J.~A., {Loch}, S.~D., {Pindzola}, M.~S., {et~al.} 2008, \pra, 78, 052708, \dodoi{10.1103/PhysRevA.78.052708}

\bibitem[{{Mao} {et~al.}(2017){Mao}, {Kaastra}, \& {Badnell}}]{mao_rrloss}
{Mao}, J., {Kaastra}, J., \& {Badnell}, N.~R. 2017, \aap, 599, A10, \dodoi{10.1051/0004-6361/201629708}

\bibitem[{{Metcalfe} {et~al.}(2023){Metcalfe}, {Buzasi}, {Huber}, {Pinsonneault}, {van Saders}, {Ayres}, {Basu}, {Drake}, {Egeland}, {Kochukhov}, {Petit}, {Saar}, {See}, {Stassun}, {Li}, {Bedding}, {Breton}, {Finley}, {Garc{\'\i}a}, {Kjeldsen}, {Nielsen}, {Ong}, {R{\o}rsted}, {Stokholm}, {Winther}, {Clark}, {Godoy-Rivera}, {Ilyin}, {Strassmeier}, {Jeffers}, {Marsden}, {Vidotto}, {Baliunas}, \& {Soon}}]{metcalfe2023}
{Metcalfe}, T.~S., {Buzasi}, D., {Huber}, D., {et~al.} 2023, \aj, 166, 167, \dodoi{10.3847/1538-3881/acf1f7}

\bibitem[{{Nikoli{\'c}} {et~al.}(2018){Nikoli{\'c}}, {Gorczyca}, {Korista}, {Chatzikos}, {Ferland}, {Guzm{\'a}n}, {van Hoof}, {Williams}, \& {Badnell}}]{nikolic2018}
{Nikoli{\'c}}, D., {Gorczyca}, T.~W., {Korista}, K.~T., {et~al.} 2018, \apjs, 237, 41, \dodoi{10.3847/1538-4365/aad3c5}

\bibitem[{{Nussbaumer} \& {Storey}(1975)}]{nussbaumer1975}
{Nussbaumer}, H., \& {Storey}, P.~J. 1975, \aap, 44, 321

\bibitem[{{Olluri} {et~al.}(2013){Olluri}, {Gudiksen}, \& {Hansteen}}]{olluri2013}
{Olluri}, K., {Gudiksen}, B.~V., \& {Hansteen}, V.~H. 2013, \apj, 767, 43, \dodoi{10.1088/0004-637X/767/1/43}

\bibitem[{{Pietarila} \& {Judge}(2004)}]{pietarila2004}
{Pietarila}, A., \& {Judge}, P.~G. 2004, \apj, 606, 1239, \dodoi{10.1086/383176}

\bibitem[{{Rao} {et~al.}(2022){Rao}, {Del Zanna}, {Mason}, \& {Dufresne}}]{rao2022}
{Rao}, Y.~K., {Del Zanna}, G., {Mason}, H.~E., \& {Dufresne}, R. 2022, \mnras, 517, 1422, \dodoi{10.1093/mnras/stac2772}

\bibitem[{{Rathore} \& {Carlsson}(2015)}]{rathore2015}
{Rathore}, B., \& {Carlsson}, M. 2015, \apj, 811, 80, \dodoi{10.1088/0004-637X/811/2/80}

\bibitem[{{Raymond} \& {Doyle}(1981)}]{raymond_doyle:1981b}
{Raymond}, J.~C., \& {Doyle}, J.~G. 1981, \apj, 247, 686, \dodoi{10.1086/159080}

\bibitem[{{Rejoub} {et~al.}(2004){Rejoub}, {Bannister}, {Havener}, {Savin}, {Verzani}, {Wang}, \& {Stancil}}]{rejoub2004}
{Rejoub}, R., {Bannister}, M.~E., {Havener}, C.~C., {et~al.} 2004, \pra, 69, 052704, \dodoi{10.1103/PhysRevA.69.052704}

\bibitem[{{Sim} \& {Jordan}(2005)}]{sim2005}
{Sim}, S.~A., \& {Jordan}, C. 2005, \mnras, 361, 1102, \dodoi{10.1111/j.1365-2966.2005.09247.x}

\bibitem[{{Spice Consortium} {et~al.}(2020){Spice Consortium}, {Anderson}, {Appourchaux}, {Auch{\`e}re}, {Aznar Cuadrado}, {Barbay}, {Baudin}, {Beardsley}, {Bocchialini}, {Borgo}, {Bruzzi}, {Buchlin}, {Burton}, {B{\"u}chel}, {Caldwell}, {Caminade}, {Carlsson}, {Curdt}, {Davenne}, {Davila}, {Deforest}, {Del Zanna}, {Drummond}, {Dubau}, {Dumesnil}, {Dunn}, {Eccleston}, {Fludra}, {Fredvik}, {Gabriel}, {Giunta}, {Gottwald}, {Griffin}, {Grundy}, {Guest}, {Gyo}, {Haberreiter}, {Hansteen}, {Harrison}, {Hassler}, {Haugan}, {Howe}, {Janvier}, {Klein}, {Koller}, {Kucera}, {Kouliche}, {Marsch}, {Marshall}, {Marshall}, {Matthews}, {McQuirk}, {Meining}, {Mercier}, {Morris}, {Morse}, {Munro}, {Parenti}, {Pastor-Santos}, {Peter}, {Pfiffner}, {Phelan}, {Philippon}, {Richards}, {Rogers}, {Sawyer}, {Schlatter}, {Schmutz}, {Sch{\"u}hle}, {Shaughnessy}, {Sidher}, {Solanki}, {Speight}, {Spescha}, {Szwec}, {Tamiatto}, {Teriaca}, {Thompson}, {Tosh}, {Tustain}, {Vial}, {Walls}, {Waltham}, {Wimmer-Schweingruber}, {Woodward}, {Young},
  {de Groof}, {Pacros}, {Williams}, \& {M{\"u}ller}}]{anderson2020}
{Spice Consortium}, {Anderson}, M., {Appourchaux}, T., {et~al.} 2020, \aap, 642, A14, \dodoi{10.1051/0004-6361/201935574}

\bibitem[{{Stancil} {et~al.}(1999{\natexlab{a}}){Stancil}, {Clarke}, {Zygelman}, \& {Cooper}}]{stancil1999si4he}
{Stancil}, P.~C., {Clarke}, N.~J., {Zygelman}, B., \& {Cooper}, D.~L. 1999{\natexlab{a}}, Journal of Physics B Atomic Molecular Physics, 32, 1523, \dodoi{10.1088/0953-4075/32/6/015}

\bibitem[{{Stancil} {et~al.}(1999{\natexlab{b}}){Stancil}, {Schultz}, {Kimura}, {Gu}, {Hirsch}, \& {Buenker}}]{stancil1999o1}
{Stancil}, P.~C., {Schultz}, D.~R., {Kimura}, M., {et~al.} 1999{\natexlab{b}}, \aaps, 140, 225, \dodoi{10.1051/aas:1999419}

\bibitem[{{Stancil} {et~al.}(2001){Stancil}, {Turner}, {Cooper}, {Schultz}, {Rakovic}, {Fritsch}, \& {Zygelman}}]{stancil2001s5}
{Stancil}, P.~C., {Turner}, A.~R., {Cooper}, D.~L., {et~al.} 2001, Journal of Physics B Atomic Molecular Physics, 34, 2481, \dodoi{10.1088/0953-4075/34/12/313}

\bibitem[{{Stancil} {et~al.}(1997{\natexlab{a}}){Stancil}, {Zygelman}, {Clarke}, \& {Cooper}}]{stancil1997n5}
{Stancil}, P.~C., {Zygelman}, B., {Clarke}, N.~J., \& {Cooper}, D.~L. 1997{\natexlab{a}}, Journal of Physics B Atomic Molecular Physics, 30, 1013, \dodoi{10.1088/0953-4075/30/4/020}

\bibitem[{{Stancil} {et~al.}(1997{\natexlab{b}}){Stancil}, {Zygelman}, {Clarke}, \& {Cooper}}]{stancil1997si5he}
---. 1997{\natexlab{b}}, \pra, 55, 1064, \dodoi{10.1103/PhysRevA.55.1064}

\bibitem[{{Stancil} {et~al.}(1998{\natexlab{a}}){Stancil}, {Havener}, {Krsti{\'c}}, {Schultz}, {Kimura}, {Gu}, {Hirsch}, {Buenker}, \& {Zygelman}}]{stancil1998c1rt}
{Stancil}, P.~C., {Havener}, C.~C., {Krsti{\'c}}, P.~S., {et~al.} 1998{\natexlab{a}}, \apj, 502, 1006, \dodoi{10.1086/305937}

\bibitem[{{Stancil} {et~al.}(1998{\natexlab{b}}){Stancil}, {Gu}, {Havener}, {Krstic}, {Schultz}, {Kimura}, {Zygelman}, {Hirsch}, {Buenker}, \& {Bannister}}]{stancil1998c1cs}
{Stancil}, P.~C., {Gu}, J.~P., {Havener}, C.~C., {et~al.} 1998{\natexlab{b}}, Journal of Physics B Atomic Molecular Physics, 31, 3647, \dodoi{10.1088/0953-4075/31/16/017}

\bibitem[{{Summers}(1974)}]{summers1974}
{Summers}, H.~P. 1974, \mnras, 169, 663, \dodoi{10.1093/mnras/169.3.663}

\bibitem[{Tayal \& Zatsarinny(2016)}]{tayal2016b}
Tayal, S., \& Zatsarinny, O. 2016, Physical Review A, 94, 042707

\bibitem[{{Tr{\"a}bert} {et~al.}(2022){Tr{\"a}bert}, {Beiersdorfer}, {Brown}, {Hell}, {Lepson}, {Fairchild}, {Hahn}, \& {Savin}}]{trabert2022}
{Tr{\"a}bert}, E., {Beiersdorfer}, P., {Brown}, G.~V., {et~al.} 2022, Atoms, 10, 115, \dodoi{10.3390/atoms10040115}

\bibitem[{{Tseng} \& {Lin}(1999)}]{tseng1999}
{Tseng}, H.~C., \& {Lin}, C.~D. 1999, Journal of Physics B Atomic Molecular Physics, 32, 5271, \dodoi{10.1088/0953-4075/32/22/305}

\bibitem[{{Vernazza} \& {Raymond}(1979{\natexlab{a}})}]{vernazza_raymond:1979}
{Vernazza}, J.~E., \& {Raymond}, J.~C. 1979{\natexlab{a}}, \apjl, 228, L89, \dodoi{10.1086/182910}

\bibitem[{{Vernazza} \& {Raymond}(1979{\natexlab{b}})}]{vernazza1979}
---. 1979{\natexlab{b}}, \apjl, 228, L89, \dodoi{10.1086/182910}

\bibitem[{{Wang} {et~al.}(2006){Wang}, {He}, {Ning}, {Liu}, {Yan}, {Stancil}, \& {Schultz}}]{wang2006si4}
{Wang}, J.~G., {He}, B., {Ning}, Y., {et~al.} 2006, \pra, 74, 052709, \dodoi{10.1103/PhysRevA.74.052709}

\bibitem[{{Wang} {et~al.}(2003){Wang}, {Stancil}, {Turner}, \& {Cooper}}]{wang2003o4}
{Wang}, J.~G., {Stancil}, P.~C., {Turner}, A.~R., \& {Cooper}, D.~L. 2003, \pra, 67, 012710, \dodoi{10.1103/PhysRevA.67.012710}

\bibitem[{{Wang} {et~al.}(2002){Wang}, {Turner}, {Cooper}, {Schultz}, {Rakovic}, {Fritsch}, {Stancil}, \& {Zygelman}}]{wang2002s5he}
{Wang}, J.~G., {Turner}, A.~R., {Cooper}, D.~L., {et~al.} 2002, Journal of Physics B Atomic Molecular Physics, 35, 3137, \dodoi{10.1088/0953-4075/35/14/310}

\bibitem[{{Wang} {et~al.}(2013){Wang}, {Zatsarinny}, \& {Bartschat}}]{wang2013ci}
{Wang}, Y., {Zatsarinny}, O., \& {Bartschat}, K. 2013, \pra, 87, 012704, \dodoi{10.1103/PhysRevA.87.012704}

\bibitem[{{Wang} {et~al.}(2014){Wang}, {Zatsarinny}, \& {Bartschat}}]{wang2014}
---. 2014, \pra, 89, 062714, \dodoi{10.1103/PhysRevA.89.062714}

\bibitem[{{Wu} {et~al.}(2009){Wu}, {Qi}, {Zou}, {Wang}, {Li}, {Buenker}, \& {Stancil}}]{wu2009o4he}
{Wu}, Y., {Qi}, Y.~Y., {Zou}, S.~Y., {et~al.} 2009, \pra, 79, 062711, \dodoi{10.1103/PhysRevA.79.062711}

\bibitem[{{Yamada} {et~al.}(1989){Yamada}, {Danjo}, {Hirayama}, {Matsumoto}, {Ohtani}, {Suzuki}, {Takayanagi}, {Tawara}, {Wakiya}, \& {Yoshino}}]{yamada1989}
{Yamada}, I., {Danjo}, A., {Hirayama}, T., {et~al.} 1989, Journal of the Physical Society of Japan, 58, 1585, \dodoi{10.1143/JPSJ.58.1585}

\bibitem[{{Yan} {et~al.}(2013){Yan}, {Wu}, {Qu}, {Wang}, \& {Buenker}}]{yan2013}
{Yan}, L.~L., {Wu}, Y., {Qu}, Y.~Z., {Wang}, J.~G., \& {Buenker}, R.~J. 2013, \pra, 88, 022706, \dodoi{10.1103/PhysRevA.88.022706}

\bibitem[{{Young}(2023)}]{2023ApJ...958...40Y}
{Young}, P.~R. 2023, \apj, 958, 40, \dodoi{10.3847/1538-4357/ad0548}

\bibitem[{{Young} {et~al.}(2018){Young}, {Keenan}, {Milligan}, \& {Peter}}]{young2018iris}
{Young}, P.~R., {Keenan}, F.~P., {Milligan}, R.~O., \& {Peter}, H. 2018, \apj, 857, 5, \dodoi{10.3847/1538-4357/aab556}

\bibitem[{{Zatsarinny} \& {Bartschat}(2012{\natexlab{a}})}]{zatsarinny2012}
{Zatsarinny}, O., \& {Bartschat}, K. 2012{\natexlab{a}}, \pra, 86, 022717, \dodoi{10.1103/PhysRevA.86.022717}

\bibitem[{{Zatsarinny} \& {Bartschat}(2012{\natexlab{b}})}]{zatsarinny2012ls}
---. 2012{\natexlab{b}}, \pra, 85, 062710, \dodoi{10.1103/PhysRevA.85.062710}

\bibitem[{{Zatsarinny} {et~al.}(2006){Zatsarinny}, {Gorczyca}, {Fu}, {Korista}, {Badnell}, \& {Savin}}]{zatsarinny2006}
{Zatsarinny}, O., {Gorczyca}, T.~W., {Fu}, J., {et~al.} 2006, \aap, 447, 379, \dodoi{10.1051/0004-6361:20053737}

\bibitem[{{Zhao} {et~al.}(2005{\natexlab{a}}){Zhao}, {Stancil}, {Gu}, {Hirsch}, {Buenker}, {Imai}, \& {Kimura}}]{zhao2005s3he}
{Zhao}, L.~B., {Stancil}, P.~C., {Gu}, J.~P., {et~al.} 2005{\natexlab{a}}, \pra, 72, 032719, \dodoi{10.1103/PhysRevA.72.032719}

\bibitem[{{Zhao} {et~al.}(2005{\natexlab{b}}){Zhao}, {Stancil}, {Gu}, {Liebermann}, {Funke}, {Buenker}, \& {Kimura}}]{zhao2005s1}
---. 2005{\natexlab{b}}, \pra, 71, 062713, \dodoi{10.1103/PhysRevA.71.062713}

\bibitem[{{Zhao} {et~al.}(2006){Zhao}, {Wang}, {Stancil}, {Gu}, {Liebermann}, {Buenker}, \& {Kimura}}]{zhao2006}
{Zhao}, L.~B., {Wang}, J.~G., {Stancil}, P.~C., {et~al.} 2006, Journal of Physics B Atomic Molecular Physics, 39, 5151, \dodoi{10.1088/0953-4075/39/24/012}

\bibitem[{{Zygelman} \& {Dalgarno}(1986)}]{zygelman1986}
{Zygelman}, B., \& {Dalgarno}, A. 1986, \pra, 33, 3853, \dodoi{10.1103/PhysRevA.33.3853}

\end{thebibliography}
\bibliographystyle{aasjournal}

\appendix

\section{A brief summary of the main IDL codes and data}

\subsection{IDL routines}

The main new code is called `\texttt{ch\_calc\_ioneq}' and is used to calculate the ion charge states for any choice of temperatures and either a fixed density or pressure. Alternatively, a grid of temperatures and related densities can be imported from a file. The subroutine `\texttt{ch\_adv\_model\_setup}' is called to import the various parameters used throughout the calculation, including fitting coefficients for the recombination rates, the list of ions included in the advanced model calculation (which is contained in a new file called `\texttt{advmodel\_list.ions}'), and the model atmosphere parameters used for calculating charge transfer rates. The advanced models are switched on by default, but can be switched off. Charge transfer is switched off by default and can be switched on by using the keyword `\texttt{ct}'. Data for a few model atmospheres have been made available for the convenience of the user. As illustrated especially in Sect.~\ref{sec:obs} above, it is strongly advised that CT is switched on when modelling any TR ions of Si. 

From this point, the ionization and recombination rates are loaded for each ion using the routine `\texttt{ch\_adv\_model\_rates}'. If available, level-resolved, direct and indirect ionization rate coefficients are stored in files ending `\texttt{.dilvl}' and `\texttt{.ealvl}', respectively, while CT ionization and recombination rate coefficients are stored in files with suffixes `\texttt{.ctilvl}' and `\texttt{.ctrlvl}', respectively. After this, level populations are solved to form overall ionization and recombination rates and then the ion balances are solved.

A few measures have been introduced to speed up the routines. The primary one is in the calculation of the overall ionization and recombination rates, which requires the relative populations of the ground and metastable levels. For the advanced models only, the number of levels included for calculating the level populations has been reduced for some ions. In doing this, it is ensured that the populations of the metastable states are not affected by more than 1\% when reducing the number of levels. The large models are mostly those that include autoionizing states, and so the optional keyword `\texttt{no\_auto}' has been implemented. This removes the autoionizing states from the level population calculation, since they are  only relevant when modelling satellite lines in the X-rays.

The resulting ion balances can be saved into a standard \chianti\ format file and/or used on-the-fly by other programs which calculate line contribution functions or intensities. Another time-saving device has been to calculate ion balances only for individual elements being modelled by the on-the-fly routines. Many existing programs have been modified to incorporate the advanced models. More details can be found in the documentation and in the headers of the programs.


\subsection{Collisional ionization data}

All the electron impact ionization data for carbon from \citet{dufresne2019} and for oxygen from \citet{dufresne2020} are incorporated into the current version, with one exception. Comparison of the ion balances from \citet{dufresne2019} and the default in \chianti\ for \ion{C}{ii} shows a significant difference at low density. This arises from the ground level ionization cross section, which is 25\% higher than experiment in \citet{dufresne2019}. For this level, \chianti\ uses the \citet{dere2007} cross section, which for this level was matched with the experiment of \cite{yamada1989}. Consequently, we retain the \citet{dere2007} cross sections for the ground level and incorporate the \citet{dufresne2019} cross sections for the metastable levels, which were in good agreement with the R-Matrix calculation of \citet{ludlow2008}.

Because of the large uncertainty in using distorted wave calculations for neutrals, \citet{dufresne2019} reduced all FAC cross sections for neutral carbon by the amount needed to bring the ground level to the experimental values at the peak. Instead of using the same method for oxygen, \citet{dufresne2020} used the CI data for ground and metastable levels from \citet{tayal2016b}, which was in excellent agreement with experiment. For the \chianti\ database, other sources of neutral ionization data were investigated. Ionization cross sections from the B-spline, \textit{R}-Matrix codes \citep{zatsarinny2006} are available from the excitation data currently used in \chianti\ for neutral carbon \citep{wang2013ci}, nitrogen \citep{wang2014}, neon \citep{zatsarinny2012ls} and magnesium \citep{barklem2017}. However, when these were checked all of the ground level cross sections were significantly below experiment. When the cross sections were converted to rate coefficients, differences of factors of three or larger at temperatures relevant for CI were found compared to the rate coefficients of \citet{dere2007}, which were adjusted to agree with experiment. It is not clear whether this is because ions in metastable levels were present in the experiment, which would cause experimental values to be overestimated, or because of uncertainties in the theoretical methods. Recent calculations and experiments for ionization of, for example, neutral Ne \citep[see][for a summary]{favreau2019} have produced yet more differences. Consequently, we retain the existing methods and data from \citet{dufresne2019} and \citet{dufresne2021picrm} for ionization rates from neutrals.

\subsection{Charge transfer data}

Details of the rates and methods used for charge transfer may be found in \citet{dufresne2021pico,dufresne2021picrm}. The same  rate coefficients have been incorporated into the current version; a summary of the sources are given below.

\subsubsection{Carbon}

Radiative and collisional CT rate coefficients between \ion{C}{i} and \ion{C}{ii} come from \cite{stancil1998c1rt} for transitions connecting ground states, while they were derived from the cross sections in \cite{stancil1998c1cs} for metastable levels. For CT from the ground of \ion{C}{iii} into \ion{C}{ii}, the \cite{errea2015} results were supplemented at low energies by the recommended cross section of \cite{janev1988}; for the metastable levels \cite{errea2000} was used. \cite{errea2015} was also used for CT from \ion{C}{iv} into \ion{C}{iii} at low energies, while \cite{tseng1999} is used for higher energies. \cite{liu2003} is used for \ion{C}{v} CT recombination with H and \cite{yan2013} for reactions involving He.

\subsubsection{Nitrogen}

\cite{lin2005} is used for ionisation and recombination between \ion{N}{i} and \ion{N}{ii}. \cite{barragan2006} is used for recombination from the ground and metastable levels of \ion{N}{iii}; CT ionisation out of \ion{N}{ii} is not relevant for the ion balance because it does not take place from the ground or metastable levels. For \ion{N}{iv}, the cross sections of \cite{bienstock1984} were supplemented at low energies by the results of \cite{gargaud1981} in order to give rates which are relevant at the formation temperature of \ion{N}{iii}. CT with He from \ion{N}{iv} was included using the rate coefficients of \cite{liu2011}. \ion{N}{v} CT with H was included from \cite{stancil1997n5}.

\subsubsection{Oxygen}

\cite{stancil1999o1} is used for transitions between the ground levels of \ion{O}{i} and \ion{O}{ii}, and \cite{kimura1997} for transitions involving the $2s^2\,2p^4\,^1D$ metastable term. \cite{barragan2006} provides the rate coefficients for CT from \ion{O}{iii} into \ion{O}{ii}. Rate coefficients for CT with H for \ion{O}{iv} to \ion{O}{iii} come from \cite{wang2003o4}; for CT with He in the same ion results of \cite{wu2009o4he} are included. Lastly, the rate coefficients provided by \cite{kingdon1996} are used for \ion{O}{v}, which are derived from the calculations of \cite{butler1980}. However, these stop at a much lower temperature than where \ion{O}{v} forms in the solar atmosphere.

\subsubsection{Neon}

The ionisation potential of \ion{Ne}{i} is reasonably close to He and so CT with He is the only important charge transfer process to consider. Radiative CT ionisation dominates at low temperatures, and the rate coefficients from \cite{liu2010ne1rad} for the ground level were incorporated. For collisional CT ionisation of \ion{Ne}{i}, data from \cite{liu2010ne1col} were supplemented with those at low energies from \cite{zygelman1986}. \cite{zhao2006} provides radiative CT with He rate coefficients for recombination from the ground and metastable levels of \ion{Ne}{iii}. \cite{imai2003} is used for collisional CT and supplemented by the values from the Okuno \& Kaneko experiment \citep[as reported by][]{imai2003} at energies below the theoretical values. \cite{rejoub2004} provides the cross sections from which rate coefficients were obtained for CT with H from \ion{Ne}{iv} into \ion{Ne}{iii}.

\subsubsection{Magnesium}

CT rate coefficients are not available for metastable levels of Mg low charge states and so this process has not been included in the advanced model for this element. The comments in Sect.~\ref{sec:ion_balances} about its importance in the solar chromosphere should be noted, however.

\subsubsection{Silicon}

For silicon, CT ionisation and recombination rate coefficients between \ion{Si}{i} and \ion{Si}{ii} in collisions with H were obtained from \cite{kimura1996si1} for both the ground and metastable terms. \cite{clarke1998} was used for CT rate coefficients between \ion{Si}{ii} and \ion{Si}{iii}. For CT with H between \ion{Si}{iii} and \ion{Si}{iv} the data come from \cite{wang2006si4}, while it comes from \cite{stancil1999si4he} for CT with He. For CT recombination with H from \ion{Si}{v} no suitable data were found for the relevant temperature range, but \cite{stancil1997si5he} was used to incorporate rate coefficients for CT with He.

\subsubsection{Sulphur}

The literature on CT involving sulphur is much less extensive than for the other elements above. The only work which includes rate coefficients for the ground and metastable levels is \cite{zhao2005s1} for reactions with H between \ion{S}{i} and \ion{Si}{ii}. Rate coefficients for CT with H from the ground of \ion{S}{iii} into \ion{Si}{ii} were derived from the cross sections of \cite{christensen1981} and \cite{bacchus1993s3}, who covered different energy regimes. CT with He for these ions is covered by \cite{zhao2005s3he}, again just from the ground term. There are no relevant studies for CT recombination with H from \ion{S}{iv}; for CT with He the rate coefficients from \cite{butler1980} were used, although these are limited in temperature and the number of states included. Finally, \cite{stancil2001s5} provide rate coefficients for \ion{S}{v} in reactions with H and \cite{wang2002s5he} for reactions with He.

\end{document}